\documentclass[preprint]{imsart}

\usepackage{amsthm,amsmath,natbib,graphicx,bm,pdflscape}
\startlocaldefs
\endlocaldefs

\begin{document}

\begin{frontmatter}

% "Title of the paper"
\title{A skew-$t$-normal multi-level reduced-rank functional PCA model with applications to
replicated `omics time series data sets}
\runtitle{A skew-$t$-normal multi-level reduced-rank FPCA model}

% indicate corresponding author with \corref{}
% \author{\fnms{John} \snm{Smith}\corref{}\ead[label=e1]{smith@foo.com}\thanksref{t1}}
% \thankstext{t1}{Thanks to somebody} 
% \address{line 1\\ line 2\\ printead{e1}}
% \affiliation{Some University}

\author{\fnms{Maurice} \snm{Berk}\ead[label=e1]{maurice.berk01@imperial.ac.uk}}
\address{Section of Paediatrics\\Department of Medicine\\Imperial College London\\Norfolk Place\\London\\W2 1PG\\\printead{e1}}
\and
\author{\fnms{Giovanni} \snm{Montana}\ead[label=e2]{giovanni.montana@imperial.ac.uk}\thanksref{T1}}
\address{Statistics Section\\Department of Mathematics\\Imperial College London\\Huxley Building\\London\\SW7 2AZ\\\printead{e2}}
\affiliation{Imperial College London}
\thankstext{T1}{We are grateful to Cheryl Hemingway and Timothy Ebbels for providing access to the
example data sets used in this article}

\runauthor{Berk and Montana}

\begin{abstract}
A powerful study design in the fields of genomics and metabolomics is the `replicated time course
experiment' where individual time series are observed for a sample of biological units, such as
human patients, termed replicates. Standard practice for analysing these data sets is to fit each
variable (e.g. gene transcript) independently with a functional mixed-effects model to account for
between-replicate variance. However, such an independence assumption is biologically implausible
given that the variables are known to be highly correlated.

In this article we present a skew-$t$-normal multi-level reduced-rank functional principal
components analysis (FPCA) model for simultaneously modelling the between-variable and
between-replicate variance. The reduced-rank FPCA model is computationally efficient and,
analogously with a standard PCA for vectorial data, provides a low dimensional representation that
can be used to identify the major patterns of temporal variation. Using an example case study
exploring the genetic response to BCG infection we demonstrate that these low dimensional
representations are eminently biologically interpretable. We also show using a simulation study that
modelling all variables simultaneously greatly reduces the estimation error compared to the
independence assumption.
\end{abstract}

%\begin{keyword}[class=AMS]
%\kwd[Primary ]{}
%\kwd{}
%\kwd[; secondary ]{}
%\end{keyword}

%\begin{keyword}
%\kwd{}
%\kwd{}
%\end{keyword}

\end{frontmatter}

\section{Introduction}

Genomics and metabolomics are two examples of a broader range of `omics domains, each of
which characterises a biological organism at a different level of biomolecular organisation. A
powerful study design within these fields is the `replicated time course experiment', in which a
sample of biological units such as human patients or laboratory rats, termed `replicates', is
studied over time in order to infer the temporal behaviour of the population as a whole. As
biological processes are inherently dynamic, these time series experiments provide greater insight
than static analyses. Data arising from these experiments presents some unique challenges compared
to more traditional time series analysis application domains. In particular, the time series are
very short with $5$ to $10$
time points being typical. This is due to the expense involved in collecting observations, not
purely in monetary terms but also due to ethical concerns about obtaining the biological samples,
and the laboratory time needed to conduct the assays. The number of replicates are equally small.
Due to the specifics of the assaying technologies utilised, the observations are
often collected with a great deal of noise, and some may simply be missing. Missing data may also
arise by design, especially when it is necessary to sacrifice replicates in animal studies in
order to obtain the biological samples. Experimental design may also lead to the time points being
irregularly spaced, in an attempt to exploit \textit{a priori} knowledge about the temporal
behaviour. Finally, the data is high dimensional with tens of thousands of variables (e.g. gene
transcripts) under study simultaneously.

In order to deal with these issues, functional data analysis (FDA) \citep{Ramsay2005} has become
a popular modelling choice in the field of genomics time series data analysis. In FDA, we assume
that our observations are noisy realisations of an underlying smooth function of time (or,
analogously, a curve) which is to be estimated. After estimation, this function is then treated
as the fundamental unit of data in any subsequent analysis, such as clustering or network
inference. Within the field of genomics, FDA approaches have been proposed for clustering
unreplicated data sets \citep{Ma2006}, detecting significant genes in
multi-sample replicated data sets \citep{Storey2005} and detecting significant genes in
cross-sectional studies \citep{Angelini2009}. We ourselves have demonstrated that such methodology
is equally well-suited to the field of metabolomics \citep{Berk2011,Montana2011}.

Despite the seeming essentiality of replication, the replicated time course study design is
rare, possibly due to a lack of appreciation that without replication inference is restricted to the
single sample under study alone, or perhaps limited by the few adequate modelling choices available.
This limited range of statistical methodology has no doubt arisen from the complexity involved in
simultaneously modelling the covariance between the variables and, for a given variable, between the
replicates. This complexity is exacerbated by the high dimensionality of the data which incurs a 
significant computational cost.

There are only a handful of approaches specifically designed for replicated genomics time series
data sets that we are aware of. The first of these, proposed by \cite{Tai2009}, does not truly
account for time as a quantitative variable in the sense that the results of an analysis would be
the same if the time points were to be permuted. Furthermore, it cannot handle missing data without
resorting to undesirable imputation procedures. The other two approaches, the functional
mixed-effects model of \cite{Storey2005}, and the functional principal components analysis (FPCA)
method proposed by \cite{Liu2009} both opt to avoid the complexity of simultaneously modelling both
levels of covariance by instead modelling each variable independently. In the case of
\cite{Storey2005}, each variable is summarised with a mean curve across all replicates. The
replicate effects are treated as scalar shifts from this mean curve, so that each replicate exhibits
exactly the same temporal profile. We have previously extended this approach to allow for more
realistic heterogeneous replicate behaviour by treating the replicate effects themselves as curves
\citep{Berk2010}. Under this model, each variable can be summarised with a mean curve and covariance
surface which describes the replicate heterogeneity. Similarly, \cite{Liu2009} use the `principal
analysis through conditional expectation' (PACE) method of \cite{Yao2005}, also summarising each
variable with a mean curve, but this time with an eigen-decomposition of the covariance surface.

All of the methods mentioned above rely on the assumption of independence between variables. It is
clear to understand the motivation for this independence assumption as, while it may be biologically
unjustified, it greatly simplifies the analysis. However, given that there
have been no proposed alternatives that do account for both levels of covariance, it has not been
possible to ascertain the true impact of this simplification, for example in terms of estimation
error. However, it is reasonable to assume, given the few observations available for each
variable, that the effect is significant.

There have been two methods proposed for accounting for multiple levels of covariance outside of
`omics application domains. The first of these, introduced by \cite{Di2009}, suggests to estimate
the covariance surface at each level using the method of moments, which is then smoothed using
thin-plate spline-smoothing. For dimensionality reduction, the resulting surfaces are subject to
eigen-decompositions. We remain sceptical, however, of the applicability of such an approach to
`omics data sets where the tiny number of time points and few replicates raises significant
concerns as to the ability of the method of moments to adequately estimate the covariance
surfaces. In contrast, \cite{Di2009} demonstrate their method on a data set with over $3,000$
replicates and $960$ time points.

The other proposed approach is the multi-level reduced-rank FPCA model of \cite{Zhou2010}, extending
the single-level reduced-rank FPCA model of \cite{James2000}. This is similar to the model that we
introduce in this article, however the key difference is that they assume the principal component
loadings at each level are normally distributed. We will demonstrate here that such an assumption is
untenable at the variable level for `omics data sets, where the small number of variables with
significantly time varying profiles, in conjunction with the high dimensionaliy of the data, leads
to distributions which exhibit a high degree of kurtosis and which may be skewed. In order to
address this issue we propose a multi-level reduced-rank FPCA model in which the variable level
loadings follow a skew-\textit{t}-normal distribution, which is a flexible four parameter
distribution allowing for both heavy tails and skewness.

\section{Methods}

\subsection{A multi-level reduced-rank FPCA model}

We assume that the observation, such as gene expression level or NMR spectrum intensity, at time
$t$ on replicate $j$ for variable $i$, $y_{ij}(t)$, is described by the following functional
mixed-effects model:
\begin{align}
\label{eqn:multilevel functional model}
y_{ij}(t) &= \mu(t) + f_i(t) + g_{ij}(t) + \epsilon_{i}(t)
\end{align}
where $\mu(t)$ is the `grand mean' across all variables and $f_i(t)$ is the offset from the grand
mean for variable $i$, so that $\mu(t) + f_i(t)$ represents the mean function for variable $i$;
$g_{ij}(t)$ is the replicate offset from the variable mean for replicate $j$, specific to variable
$i$; $\epsilon_{i}(t)$ is an error term, specific to variable $i$. Note that if the replicate effect
$g_{ij}(t)$ was not variable specific then the subscript $i$ could be dropped so that the same
$g_{j}(t)$ term was shared across all variables. However, both intuition and real data sets support
the idea of separate replicate effects for each variable. We would expect in a genomics experiment,
for instance, that certain gene transcripts display a homogeneous response across all replicates
while others are much more heterogeneous, and it could well be the case that these differences lead
to exactly the biological effect we are seeking to identify. Even when two transcripts both display
heterogeneity between the replicates, the exact nature of that variation is likely to be
transcript-specific. In Supplementary Figures 1 and 2 we give raw data for two transcripts from our
example data set that illustrate these points.

As in \cite{James2000}, we can simultaneously achieve computational efficiency, parsimony and
dimensionality reduction by replacing $f_{i}(t)$ and $g_{ij}(t)$ in
(\ref{eqn:multilevel functional model}) with their Karhunen-Lo\`{e}ve decompositons, yielding
\begin{align*}
y_{ij}(t) &= \mu(t) + \sum_{k=1}^{\infty} \zeta_{k}(t) \alpha_{ik} + \sum_{l=1}^{\infty}
\eta_{il}(t)\beta_{ijl} + \epsilon_{ij}(t)
\end{align*} 
where $\zeta_{k}(t)$ is the $k$-th principal component function at the variable level, $\alpha_{ik}$
is variable $i$'s loading on the $k$-th principal component function, $\eta_{il}(t)$ is the $l$-th
principal component function at the replicate level, specific to variable $i$ and $\beta_{ijl}$ is
replicate $j$'s loading on the $l$-th principal component function specific to variable $i$.
Truncating the decompositions at the $K$-th and $L_i$-th component for the variable and replicate
level respectively gives
\begin{align*}
y_{ij}(t) &= \mu(t) + \sum_{k=1}^{K} \zeta_{k}(t) \alpha_{ik} + \sum_{l=1}^{L_i}
\eta_{il}(t)\beta_{ijl} + \epsilon_{ij}(t)
\end{align*}
Note that the number of principal components retained at the replicate level, $L_i$, is variable
specific as indicated by the subscript. The optimal number of principal components to be retained at
this level will differ between variables depending upon the amount of between-replicate
heterogeneity displayed. Representing the functions $\mu(t)$, $\zeta_k(t)$ and $\eta_{il}(t)$ using
an appropriately orthogonalised $p$-dimensional B-spline basis \citep{Zhou2008} and collecting all
$N_{ij}$ observations on replicate $j$ for variable $i$ in the vector $\bm{y}_{ij}$ yields
\begin{align*}
\bm{y}_{ij} &= \bm{B}_{ij}\bm{\theta}_\mu + \sum_{k=1}^{K}\bm{B}_{ij}\bm{\theta}_{\alpha_k}
\alpha_{ik} + \sum_{l=1}^{L_i}\bm{B}_{ij}\bm{\theta}_{\beta_{il}}\beta_{ijl} + \bm{\epsilon}_{ij}
\end{align*}
where $\bm{B}_{ij}$ is the $N_{ij} \times p$ B-spline basis matrix that has been transformed such
that $(L/g)\bm{B}^{T}\bm{B} = \bm{I}$ where $\bm{B}$ is the basis evaluated on a fine grid of points
covering the range of the time course, $L$ is the length of this fine grid, $g$ is the distance
between successive grid points, $\bm{\theta}_{\mu}$ is a $p$-length vector of fitted spline
coefficients for the grand mean function $\mu(t)$, $\bm{\theta}_{\alpha_k}$ is a $p$-length vector
of fitted spline coefficients for the $k$-th principal component function at the variable level and
$\bm{\theta}_{\beta_{il}}$ is a $p$-length vector of fitted spline coefficients for the $l$-th
principal component function at the replicate level. The transformation of $\bm{B}$ is required to
enforce the FPCA orthogonality constraint that
\begin{equation*}
\int_t \zeta_{k}(t)\zeta_{k'}(t) = \left\{ \begin{array}{ll} 1 \quad & k = k'  \\ 0 \quad & k \ne k'
\end{array} \right.
\end{equation*}
and similarly for $\eta_{il}(t)$. 

Defining the $p \times K$ matrix
$\bm{\Theta}_{\alpha} = [\bm{\theta}_{\alpha_1} \cdots \bm{\theta}_{\alpha_K}]$ and the
$p \times L_i$ matrix $\bm{\Theta}_{\beta_{ij}} = [\bm{\theta}_{\beta_{i1}} \cdots
\bm{\theta}_{\beta_{iL_i}}]$ allows the summations to be simplified using matrix algebra as
\begin{align}
\bm{y}_{ij} &= \bm{B}_{ij}\bm{\theta}_\mu + \bm{B}_{ij}\bm{\Theta}_{\alpha_k}\bm{\alpha}_{i} +
\bm{B}_{ij}\bm{\Theta}_{\beta_{i}}\bm{\beta}_{ij} + \bm{\epsilon}_{ij}
\label{eqn:multi-level reduced-rank fpca model}
\end{align}
where $\bm{\alpha}_{i} = [\alpha_{i1} \cdots \alpha_{iK}^{T}]$ is the $K$-length vector formed by
collecting all of the $\alpha_{ik}$ terms together and similarly for $\bm{\beta}_{ij}$. By
collecting the observations on all replicates $j = 1,\cdots,n_{i}$ for variable $i$, in the
vector $\bm{y}_{i} = [\bm{y}_{i1} \cdots \bm{y}_{in_{i}}]^{T}$, we can write
\begin{align*}
\bm{y}_{i} &= \bm{B}_{i}\bm{\theta}_{\mu} + \bm{B}_{i}\bm{\Theta}_{\alpha}\bm{\alpha}_{i} +
\widetilde{\bm{B}_{i}}\widetilde{\bm{\Theta}_{\beta_i}}\bm{\beta}_{i} + \bm{\epsilon}_{i}
\end{align*}
where $\bm{B}_{i} = [\bm{B}_{i1}^{T} \cdots \bm{B}_{in_{i}}^{T}]^{T}$ is the $N_{i} \times p$ basis
matrix formed by stacking each $\bm{B}_{ij}$ matrix on top of one another. There are $n_{i}$ such
matrices, each with $N_{ij}$ rows and so the matrix $\bm{B}_{i}$ has $N_{i}$ rows where $N_{i} =
\sum_{i=1}^{n_i} N_{ij}$ is the total number of observations on variable $i$ across all replicates.
In contrast, the matrix $\widetilde{\bm{B}_{i}} = \mbox{diag}(\bm{B}_{i1},\cdots,\bm{B}_{in_i})$ is
a block diagonal matrix of dimension $N_{i} \times (n_{i}p)$ where the blocks correspond to the
$\bm{B}_{ij}$ matrices. Similarly, $\widetilde{\bm{\Theta}_{\beta_{i}}} =
\mbox{diag}(\bm{\Theta}_{\beta_{i}},
\cdots,\bm{\Theta}_{\beta_{i}})$ is a block diagonal matrix of dimension $(n_{i}L_{i}) \times 
(n_{i}L_{i})$ where each block is identical and equal to $\bm{\Theta}_{\beta_{i}}$. Finally,
$\bm{\beta}_{i} = [\bm{\beta}_{i1} \cdots \bm{\beta}_{in_{i}}]^{T}$ and
$\bm{\epsilon}_{i} = [\bm{\epsilon}_{i1}\cdots \bm{\epsilon}_{in_i}]^{T}$.

Standard practice would be to assume that $\bm{\alpha}_{i}$, $\bm{\beta}_{ij}$ and
$\bm{\epsilon}_{ij}$ are all independently multivariate normally distributed with zero mean and
covariance matrices $\bm{D}_{\alpha}$, $\bm{D}_{\beta_i}$ and $\sigma_{i}^{2}\bm{I}$ respectively.
Under these assumptions, $\bm{y}_{i}$ is marginally multivariate normal and the model parameters can
be estimated by treating the principal component loadings $\bm{\alpha}_{i}$ and $\bm{\beta}_{ij}$ as
missing data and employing the EM algorithm. Technical details for this approach can be found in
the Supplementary Material.

\subsection{A skew-$t$-normal multi-level reduced-rank FPCA model}

\begin{figure}[tbp]
\includegraphics[width=0.5\textwidth]{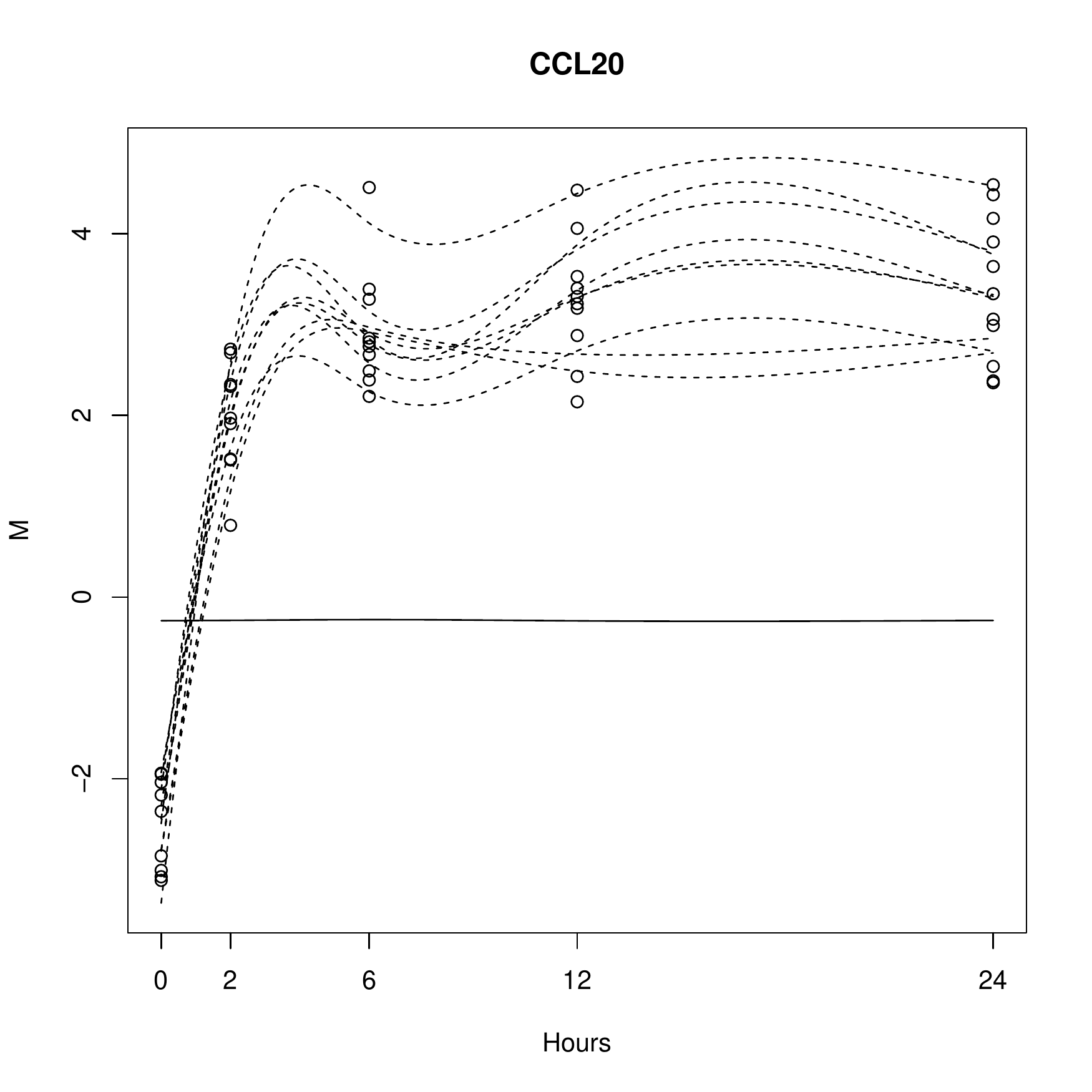}\includegraphics[width=0.5\textwidth]{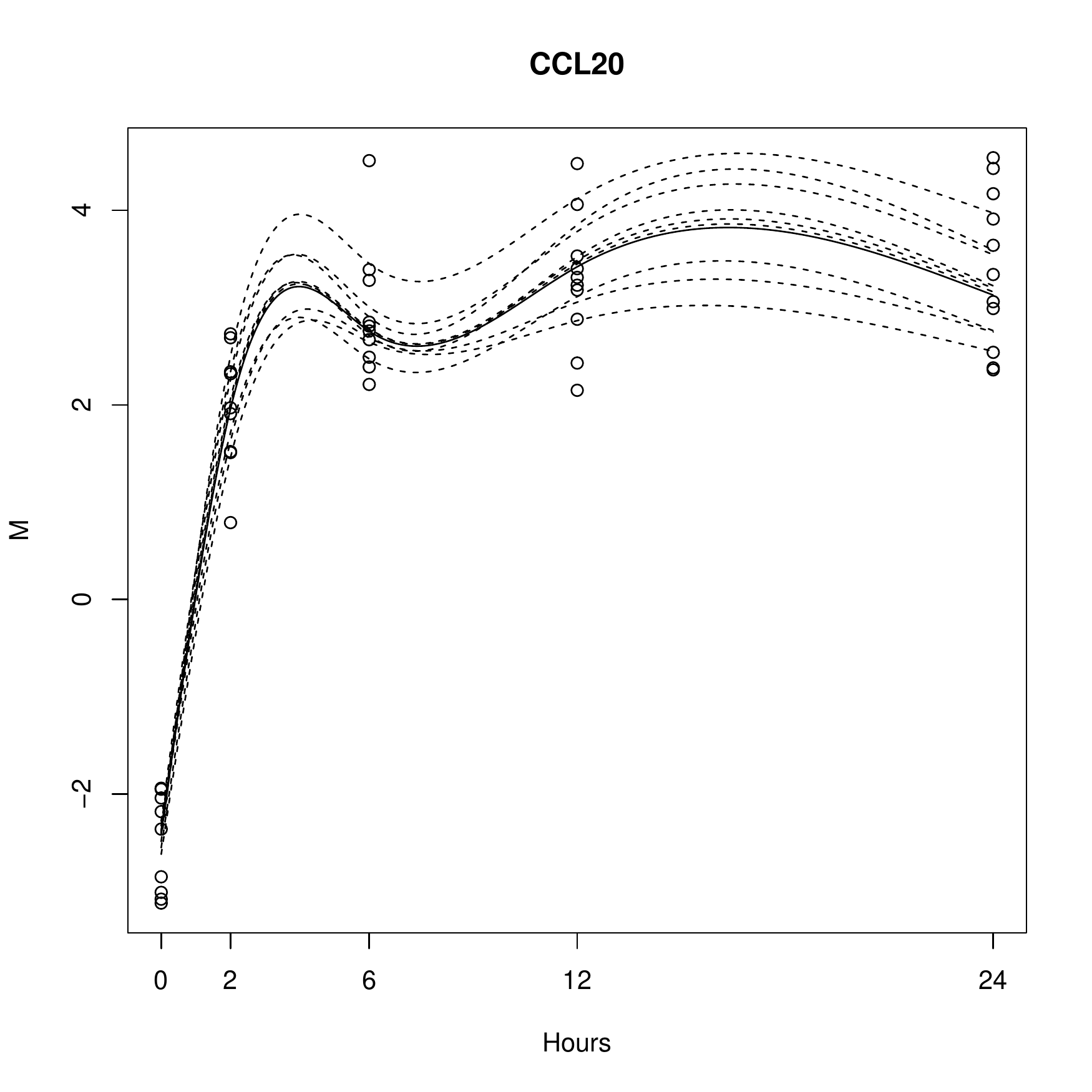}
\caption
{\label{fig:gaussian problem}Left: An example of the poor model fits obtained under the
Gaussian reduced-rank multi-level FPCA model with real data. This example gene transcript is typical
of variables across a range of data sets, where, while the replicate curves, indicated by the dashed
lines, look good and closely map the underlying observations, the mean curve, given by the solid
line, seems to be biologically implausible. Right: This poor fit is fixed when we instead adopt our
proposed skew-\textit{t}-normal multi-level reduced-rank FPCA model.}
\end{figure}

We discovered upon fitting the multi-level reduced-rank FPCA model with the normal assumption to
real data that the fits were biologically implausible, an example of which is given on the left hand
side of Figure \ref{fig:gaussian problem}. As can be clearly seen, the mean curve for this gene
transcript, indicated by the solid line, does an unusually poor job of describing the underlying
observations, being flat over the entire range of the time course. However, the replicate-level
curves, indicated by the dashed lines, follow the observations much more closely. Upon further
investigation, it became clear that this was due to an extreme departure from normality for the
principal component loadings at the variable level, as can be seen in histograms of the initalised
loadings for two example data sets given in Supplementary Figures 3 and 4, which exhibit heavy
tails and varying degrees of skewness. To deal with this issue, we propose to instead adopt the
assumption that the variable level loadings follow a skew-$t$-normal distribution, which is flexible
enough to account for the heterogenous departures from normality. Formally, we assume that:
\begin{equation}
\begin{gathered}
\label{eqn:skew-t-normal mlfpca assumptions}
\alpha_{ik} \stackrel{\mathrm{i.i.d.}}{\sim} StN(\xi_{\alpha_k}, \sigma_{\alpha_{k}}^{2},
\lambda_{\alpha_{k}},
\nu_{\alpha_{k}}) \\
\alpha_{ik} \perp \alpha_{ik'}, k \ne k' \quad \alpha_{ik} \perp \alpha_{i'k}, i \ne i'\\
E[\alpha_{ik}] = 0
\end{gathered}
\end{equation}
while retaining the assumption that $\bm{\beta}_{ij}$ and $\bm{\epsilon}_{ij}$ are multivariate
normally distributed. $z \sim StN(\xi, \sigma^2, \lambda, \nu)$ denotes that the random variable $z$
follows a skew-$t$-normal distribution \citep{G'omez2007} where $\xi$ is a location parameter,
$\sigma^2$ is a scale parameter, $\lambda$ is a skewness parameter and $\nu$ is the degrees of
freedom controlling the kurtosis. The density of $z$ is given by
\begin{align}\label{eqn:skew-t-normal density}
f(z|\xi,\sigma^2,\lambda,\nu) = 2t_{\nu}(z;\xi,\sigma^2)\Phi\left(\frac{z - \xi}{\sigma}\lambda
\right)
\end{align}
where $t_{\nu}(z;\xi,\sigma^2)$ denotes the Student-\textit{t} density with $\nu$ degrees of
freedom, location parameter $\xi$ and scale parameter $\sigma^2$, and $\Phi$ denotes the normal
cumulative distribution function. We note here in passing the closely related skew-\textit{t}
distribution of \cite{Azzalini2003} which is identical in form to (\ref{eqn:skew-t-normal density})
except that the the normal cumulative distribution function, which controls the skewness of the
density, is replaced by the Student-\textit{t} cumulative distribution function, and depends not
just on the skewness parameter $\lambda$ but also the degrees of freedom $\nu$. As \cite{Ho2010}
point out, evaluating the skew-\textit{t}-normal density is therefore computationally simpler and
the decoupling of the skewness from the degrees of freedom parameter is more conceptually sound.

Note that we have chosen not to use a multivariate skew-\textit{t}-normal density for the
variable-level loadings as this would require a single degrees of freedom parameter to be shared
across all components, and we assume that they are independent anyway for the purposes of
identifiability. Furthermore, we retain the original assumption that the replicate-level loadings
follow a multivariate normal distribution. Judging from Figure \ref{fig:gaussian problem}, which is
typical of other variables across multiple data sets, the replicate-level curves remain plausible
under this assumption.

As in the Gaussian case, we can estimate the parameters under the assumptions given in
(\ref{eqn:skew-t-normal mlfpca assumptions}) by treating the principal component loadings as missing
data and employing the EM algorithm. The maximum likelihood estimators of $\bm{\theta}_{\mu}$,
$\bm{\Theta}_{\alpha}$, $\bm{\Theta}_{\beta_i}$, $\bm{D}_{\beta_i}$ and $\sigma_{i}^{2}$ can be
derived analytically as illustrated in the Supplementary Material. For the parameters of the
skew-$t$-normal distributions, these can be estimated using Newton-Raphson (NR) as in
\cite{G'omez2007} or the EM algorithm as in \cite{Ho2010}. However, the NR approach relies on
numerical integration and the EM algorithm is known to be slow to converge; alternatively, we have
found that a simplex optimisation \citep{Nelder1965} works very well in practice.

The conditional expectations that need to be calculated at the E-step of the EM algorithm based on
these MLE estimators are summarised in Supplementary Table 2. However, expressions for these are
challenging
to obtain given that under the distributional assumptions, the marginal density of $\bm{y}_{i}$, as
a sum of skew-\textit{t}-normal and normal distributed random variables, follows no known form. In
these circumstances we can instead turn to the \textit{Monte Carlo} (MC) EM-algorithm
\citep{Wei1990} which replaces the calculation of intractable expectations at the E-step with
approximations based on averaging random samples drawn from the target density. As the target
density is itself unknown, it is necessary to resort to stochastic simulation algorithms.

Given that the skew-$t$-normal distribution admits a convenient hierarchical representation, we
suggest the use of the Gibbs sampler for drawing samples from the joint conditional distribution
$f(\bm{\alpha}_{i},\bm{\beta}_{ij}|\bm{y}_{i})$. Specifically, following \cite{Ho2010}, if
$\tau_{ik} \sim \Gamma(\nu_{\alpha_k}/2,\nu_{\alpha_k}/2)$ then
\begin{equation}
\begin{gathered}
\gamma_{ik} | \tau_{ik} \sim TN\left(0,\displaystyle
 \frac{\tau_{ik} + \lambda_{\alpha_k}^{2}}{\tau_{ik}};
(0,\infty)\right) \\
\alpha_{ik}|\gamma_{ik},\tau_{ik} \sim N\left( \displaystyle 
\xi_{\alpha_k} + \frac{\sigma_{\alpha_k}\lambda_{\alpha_k}}
{\tau_{ik} + \lambda_{\alpha_k}^{2}}\gamma_{\alpha_k},\frac{\sigma_{\alpha_k}^{2}}{\tau_{ik} +
\lambda_{\alpha_k}^{2}}\right)
\end{gathered}
\label{eqn:hierarchical model skew-t-normal}
\end{equation}
where $TN(\mu,\sigma^{2};(a,b))$ denotes the truncated normal distribution lying within the interval
$(a,b)$. It can then be shown (see Supplementary Material) that
\begin{gather}
\label{eqn:gamma conditional}
\gamma_{ik}|\alpha_{ik} \sim TN(
(\alpha_{ik} - \xi_{\alpha_k})\frac{\lambda_{\alpha_k}}{\sigma_{\alpha_k}},1;(0,\infty)) \\
\label{eqn:tau conditional}
\tau_{ik}|\alpha_{ik} \sim \Gamma \left(\frac{\nu_{\alpha_k}+1}{2},
\frac{\nu_{\alpha_k} + (\alpha_{ik} - \xi_{\alpha_k})^{2}/\sigma_{\alpha_k}^{2}}{2}\right)
\end{gather}
Therefore we introduce additional latent variables,
$\bm{\tau}_{i} = [\tau_{i1} \cdots \tau_{iK}]^{T}$ and
$\bm{\gamma}_{i} = [\gamma_{i1} \cdots \gamma_{iK}]^{T}$, and the target density for the Monte
Carlo E-step becomes the joint conditional distribution $f(\bm{\alpha}_{i},\bm{\beta}_{i},
\bm{\tau}_{i},\bm{\gamma}_{i}|\bm{y}_{i})$ whose form is still unknown. However, the conditionals
$f(\bm{\alpha}_{i},\bm{\beta}_{i}|\bm{y}_{i},\bm{\tau}_{i},\bm{\gamma}_{i})$,
$f(\bm{\tau}_{i}|\bm{y}_{i},\bm{\alpha}_{i},\bm{\beta}_{i},\bm{\gamma}_{i})$ and
$f(\bm{\gamma}_{i}|\bm{y}_{i},\bm{\alpha}_{i},\bm{\beta}_{i},\bm{\tau}_{i})$ follow known
distributions that are easy to sample from and hence the Gibbs sampler can be used to efficiently
generate samples that are approximately distributed according to the target full joint conditional
density.

Starting with $f(\bm{\alpha}_{i},\bm{\beta}_{i}|\bm{y}_{i},\bm{\tau}_{i},\bm{\gamma}_{i})$ note
that from (\ref{eqn:hierarchical model skew-t-normal}), conditional on
$\tau_{ik}$ and $\gamma_{ik}$, $\alpha_{ik}$ is normally distributed with mean $\mu_{\alpha_k} =
\xi_{\alpha_k} + \frac{\sigma_{\alpha_k}\lambda_{\alpha_k}}{\tau_{ik} + \lambda_{\alpha_k}^{2}}
\gamma_{\alpha_k}$ and variance $v_{\alpha_k} = \frac{\sigma_{\alpha_k}^{2}}
{\tau_{ik} + \lambda_{\alpha_k}^{2}}$. Hence we can write
\begin{multline*}
\left.\left[\begin{array}{c}\bm{\alpha}_{i} \\ \bm{\beta}_{i} \\ \bm{y}_{i} \end{array}\right]
\right|\bm{\tau}_{i},\bm{\gamma}_{i} \sim MVN\left(\left[\begin{array}{c}\bm{\mu}_{\alpha} \\ \bm{0}
\\ \bm{B}_{i}\bm{\theta}_{\mu} + \bm{B}_{i}\bm{\Theta}_{\alpha}\bm{\mu}_{\alpha}
\end{array}\right]\right.,
\\
\left.\left[\begin{array}{ccc}\mbox{diag}(\bm{v}_{\alpha}) & \bm{0} & 
\mbox{diag}(\bm{v}_{\alpha})\bm{\Theta}_{\alpha}^{T}\bm{B}_{i}^{T}\\ \bm{0} &
\widetilde{\bm{D}_{\beta_i}} & \widetilde{\bm{D}_{\beta_i}}\widetilde{\bm{\Theta}_{\beta_i}^{T}}
\widetilde{\bm{B}_{i}^{T}} \\
\bm{B}_{i}\bm{\Theta}_{\alpha}\mbox{diag}(\bm{v}_{\alpha}) & \widetilde{\bm{B}_{i}}
\widetilde{\bm{\Theta}_{\beta_i}}\widetilde{\bm{D}_{\beta_i}} & \bm{V}_{y_i|\tau_i,\gamma_i}
\end{array}\right]\right)
\end{multline*}
where $\bm{\mu}_{\alpha} = [\mu_{\alpha_1} \cdots \mu_{\alpha_K}]^{T}$ and $\bm{v}_{\alpha} =
[v_{\alpha_1} \cdots v_{\alpha_K}]^{T}$, $\bm{V}_{y_i|\tau_i,\gamma_i} = 
\sigma^2_{i}\bm{I}_{N_i \times N_i} + 
\bm{B}_{i}\bm{\Theta}_{\alpha}\mbox{diag}(\bm{v}_{\alpha})\bm{\Theta}_{\alpha}^{T}\bm{B}_{i}^{T} +
\widetilde{\bm{B}_{i}}
\widetilde{\bm{\Theta}_{\beta_i}}\widetilde{\bm{D}_{\beta_i}}\widetilde{\bm{\Theta}_{\beta_i}^{T}}
\widetilde{\bm{B}_{i}^{T}}$ and
$\widetilde{\bm{D}_{\beta_i}} = \mbox{diag}(\bm{D}_{\beta_i},\cdots,\bm{D}_{\beta_i})$. Using a
standard result from multivariate statistics \citep{Anderson1958}, we have that
$f(\bm{\alpha}_{i},\bm{\beta}_{i}|\bm{y}_{i},\bm{\tau}_{i},\bm{\gamma}_{i})$ is therefore also
multivariate normal (see Supplementary Material for specification of the mean and covariance), which
is simple to sample from. For
$f(\bm{\gamma}_{i}|\bm{y}_{i},\bm{\alpha}_{i},\bm{\beta}_{i},\bm{\tau}_{i})$ and
$f(\tau_{i}|\bm{y}_{i},\bm{\alpha}_{i},\bm{\beta}_{i},\bm{\gamma}_{i})$, we use
(\ref{eqn:gamma conditional}) and (\ref{eqn:tau conditional}) respectively.

\subsection{MCEM Algorithm Summary}

Having derived a process for calculating the required maximum likelihood estimators and conditional
expectations, we are now in a position to summarise the complete MCEM algorithm for estimating the
model parameters for the skew-$t$-normal multi-level reduced-rank FPCA model.

A procedure for initialisation is described in the Supplementary Material.  After initialisation,
the algorithm alternates between approximating the conditional expectations by
running the Gibbs sampler for each variable at the E-step, and calculating the maximum likelihood
estimators with their sufficient statistics replaced by the conditional expectations. Note that the
maximum likelihood estimators of $\bm{\theta}_{\mu}$, and the individual columns of
$\bm{\Theta}_{\alpha}$ and $\bm{\Theta}_{\beta_i}$ are interdependent. Therefore it is necessary to
employ an Expectation \textit{Conditional} Maximisation (ECM) algorithm where each part of the
M-step is carried out holding all of the other parameters fixed. In other words, first
$\bm{\theta}_{\mu}$ is estimated holding $\bm{\Theta}_{\alpha}$ and $\bm{\Theta}_{\beta_i}$ fixed.
Next, the first column of $\bm{\Theta}_{\alpha}$ is estimated, holding all other columns of
$\bm{\Theta}_{\alpha}$ fixed along with $\bm{\Theta}_{\beta_i}$ and $\bm{\theta}_{\mu}$, and so on.
This process can be iterated as in \cite{James2000} or performed once per M-step as in
\cite{Zhou2010}. We have found the latter to work better in practice, in terms of numerical
stability.

The maximum likelihood estimates of $\bm{\Theta}_{\alpha}$ and $\bm{\Theta}_{\beta_i}$ are not
guaranteed to satisfy the constraint that the columns are orthogonal. While it is trivial to
orthogonalise them by carrying out an eigen-decomposition, there is a lack of consensus as to whether
this should be carried out at every iteration as in \cite{Zhou2008,Zhou2010} or once the EM
algorithm has converged as in \cite{James2000}. In the case of the former we sacrifice
the monotonicity property of the EM algorithm in order to ensure that the estimates remain within
the valid parameter space at each iteration. In the case of the latter the monotonicity property is
retained but intuition suggests that the algorithm may converge to parameter values that are far
from those which maximise the constrained likelihood once the orthogonalisation is carried out. In
fact, in our experience with the model under the assumption of normality, orthogonalising at each
iteration results in the algorithm converging to a marginally larger likelihood at the expense of
many more iterations required before convergence, possibly due to some of the iterations decreasing
the likelihood. Furthermore, we have found that when the full-rank model is fit, such that $K$ and
$L_i$ are set to the maximum permitted by the choice of spline basis, computational
instability can occur when orthogonalising at each iteration. An alternative to either of these
approaches is to carry out the maximisation within the constrained parameter space - i.e. all
matrices with orthonormal columns - as in \cite{Peng2009}. Such methods are difficult to implement
due to their mathematical complexity and the success of \cite{James2000,Zhou2008,Zhou2010} suggest
that, practically speaking, they are unnecessary.

The complete algorithm is as follows:
\begin{enumerate}
\item Initialise parameters as described in the Supplementary Material
\item E-step: For each variable, run the Gibbs sampler for $S$ iterations, sampling from
$f(\bm{\alpha}_{i},\bm{\beta}_{i}|\bm{y}_{i},\bm{\tau}_{i},\bm{\gamma}_{i})$,
$f(\bm{\tau}_{i}|\bm{y}_{i},\bm{\alpha}_{i},\bm{\beta}_{i},\bm{\gamma}_{i})$ and
$f(\bm{\gamma}_{i}|\bm{y}_{i},\bm{\alpha}_{i},\bm{\beta}_{i},\bm{\tau}_{i})$ in turn
\item M-step Step 1: Find the maximum likelihood estimators of the parameters of the
skew-$t$-normal distributions using a simplex optimisation
\item M-step Step 2: Update $\widehat{\bm{D}_{\beta_i}}$, $i=1,\cdots,M$
\item M-step Step 3: Update $\widehat{\sigma_{i}^{2}}$, $i=1,\cdots,M$
\item M-step Constrained Maximisation Step 1: Update $\widehat{\bm{\theta}}_{\mu}$ while holding
$\widehat{\bm{\Theta}_{\alpha}}$ and $\widehat{\bm{\Theta}_{\beta_i}}$, $i=1,\cdots,M$ fixed
\item M-step Constrained Maximisation Step 2: Update $\widehat{\bm{\Theta}_{\alpha}}$ while holding
$\widehat{\bm{\theta}}_{\mu}$ and $\widehat{\bm{\Theta}_{\beta_i}}$, $i=1,\cdots,M$ fixed
\item M-step Constrained Maximisation Step 3: Update $\widehat{\bm{\Theta}_{\beta_i}}$,
$i=1,\cdots,M$ while holding $\widehat{\bm{\theta}}_{\mu}$ and $\widehat{\bm{\Theta}_{\alpha}}$
fixed
\item Check for convergence. If not converged, return to  3.
\item Orthogonalise $\widehat{\bm{\Theta}_{\alpha}}$ and $\widehat{\bm{\Theta}_{\beta_i}}$,
$i=1,\cdots,M$
\end{enumerate}

\subsection{Model Selection}

Two remaining issues are how to select the number of principal components, both at the variable-
and replicate-level, and what spline basis to use. For selecting the number of principal components,
two main approaches can be considered. In the first, the proportion of variance explained by each
principal component function can be approximated by fitting the full-rank model - such that $K$ and
$L_i$ (for all $i$) are the maximum permitted by the number of design time points \citep{James2000}.
For the second method, cross-validation is used to score each potential value of $K$ and $L_i$
\citep{Zhou2010}. Note that both of these approaches require a subjective
interpretation. For the proportion of variance explained method, either an arbitrary cutoff for
cumulative variance such as $95\%$ or $99\%$ must be used, perhaps aided by an examination of a
scree plot and a visualisation of the principal component functions in order to ascertain their
interpretability. On the other hand, as \cite{Zhou2010} discuss, when using the cross-validation
method on real data the score may simply decrease as the number of principal components increases,
which results in always selecting the full-rank model going by the score alone. Therefore they
suggest to instead subjectively trade-off between parsimony and cross-validation score, again with
the aid of a scree plot. However, in the model of \cite{Zhou2010}, the number of principal
components at the second-level is not dependent on the first-level and so the
cross-validation method is much more tractable in their
setting than ours, where the high-dimensional optimisation renders the approach impractical. By
necessity therefore, we employ the proportion of variance explained method. On simulated data we
have discovered that this works very well at identifying the correct number of principal components
at the variable-level, but struggles at the replicate-level, most likely due to the much smaller
sample sizes. We therefore suggest that a (much) more conservative criteria be applied at the
replicate-level. For instance, we have found that retaining those principal components that explain
$99\%$ of the variance at the variable-level and $60\%$ of the variance at the replicate-level
worked well on these data sets.

For selecting the spline basis we suggest to use natural cubic splines with a knot placed at each
design time point for several reasons. Firstly, this avoids the computational burden of having to
select the number of knots. Secondly, many of the data sets provided by our collaborators have
unequally spaced time points, and it therefore makes sense to use each time point as a knot in order
to adequately capture the temporal dynamics. Thirdly, the reason to countenance against such an
approach would be the danger of overfitting. This is accounted for through the use of the
reduced-rank model, essentially placing a rank-constraint on the covariance matrix of the spline
basis coefficients. Conceptually speaking, this approach to spline basis selection is quite similar
to the use of smoothing splines, where a knot is placed at each design time point and overfitting is
avoided through the use of a penalty parameter on the likelihood.

\section{Results}

\subsection{Simulation comparison of the Gaussian single- and multi-level reduced-rank FPCA models}
\label{sec:gaussian mlfpca simulation}
We set out to determine whether single-level approaches that assume the variables are independent
are adequate when it comes to estimating the true underlying curves, and to quantify the improvement
that can be gained through the use of a multi-level model using the following simulation setting.

We generated data under the multi-level reduced-rank FPCA model
(\ref{eqn:multi-level reduced-rank fpca model}) with normality assumed, as the skew-$t$-normal model
is too computationally intensive, with its reliance on MC methods, for large scale simulation
studies. We fixed the number of principal components at the variable level to $K=2$ with a single
principal component at the replicate level for each variable, so that $L_i=1$ for all $i$. We used a
B-spline basis with a single knot placed at the centre of the time course. The spline coefficients
for the grand mean, $\bm{\theta}_{\mu}$, were fixed to produce the curve that can be seen in
Figure \ref{fig:mlfpca simulation mean plus pcs}.

\begin{figure}[p]
\includegraphics[width=\textwidth]{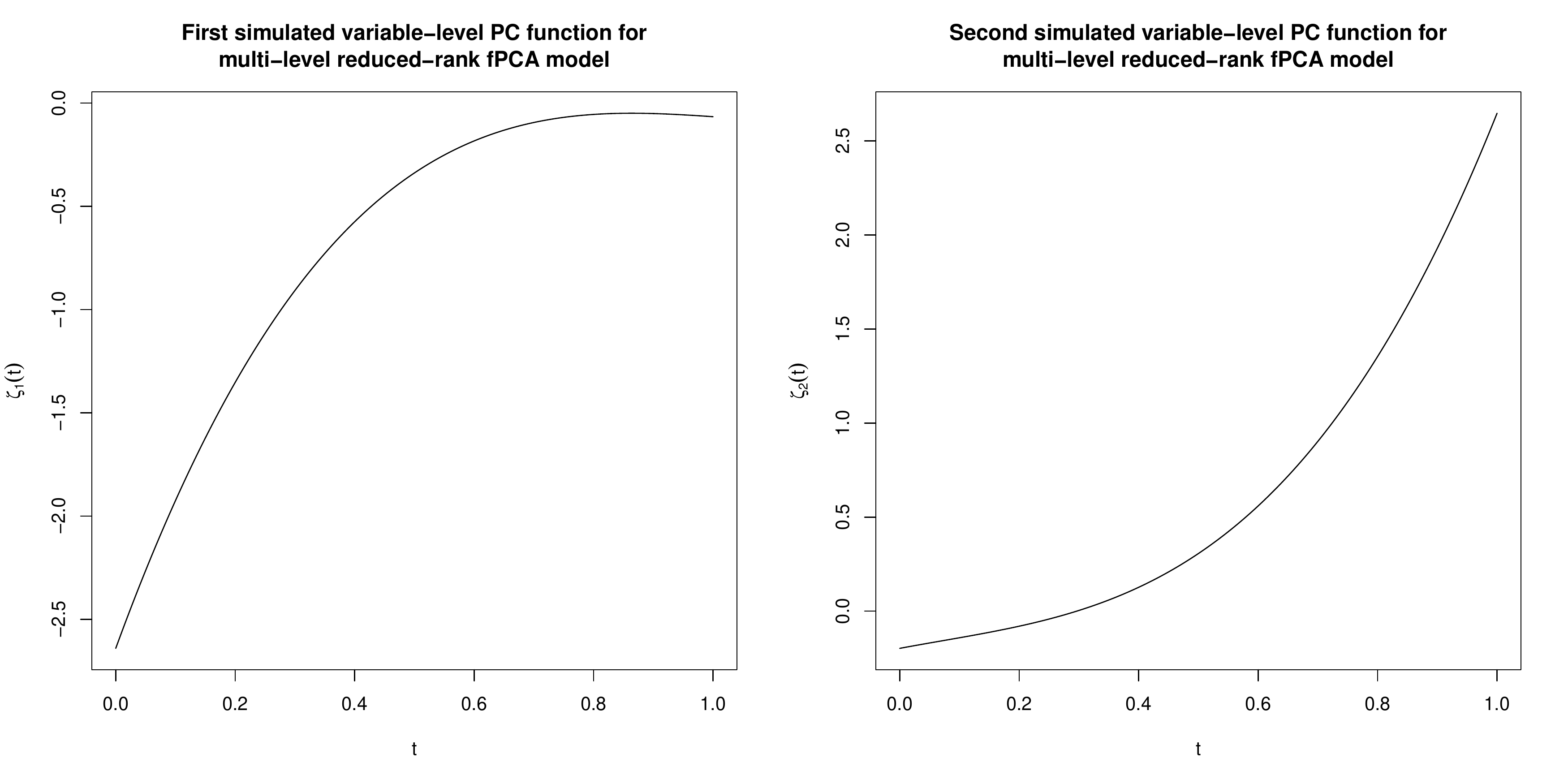}
\caption
{\label{fig:mlfpca simulation pcs}Two simulated variable-level principal component functions
used in our simulation study to compare the Gaussian single- and multi-level reduced-rank FPCA
models. The
principal component functions were set to explain $75\%$ and $25\%$ of the variance in the data
respectively.}
\end{figure}

\begin{figure}[p]
\includegraphics[width=\textwidth]{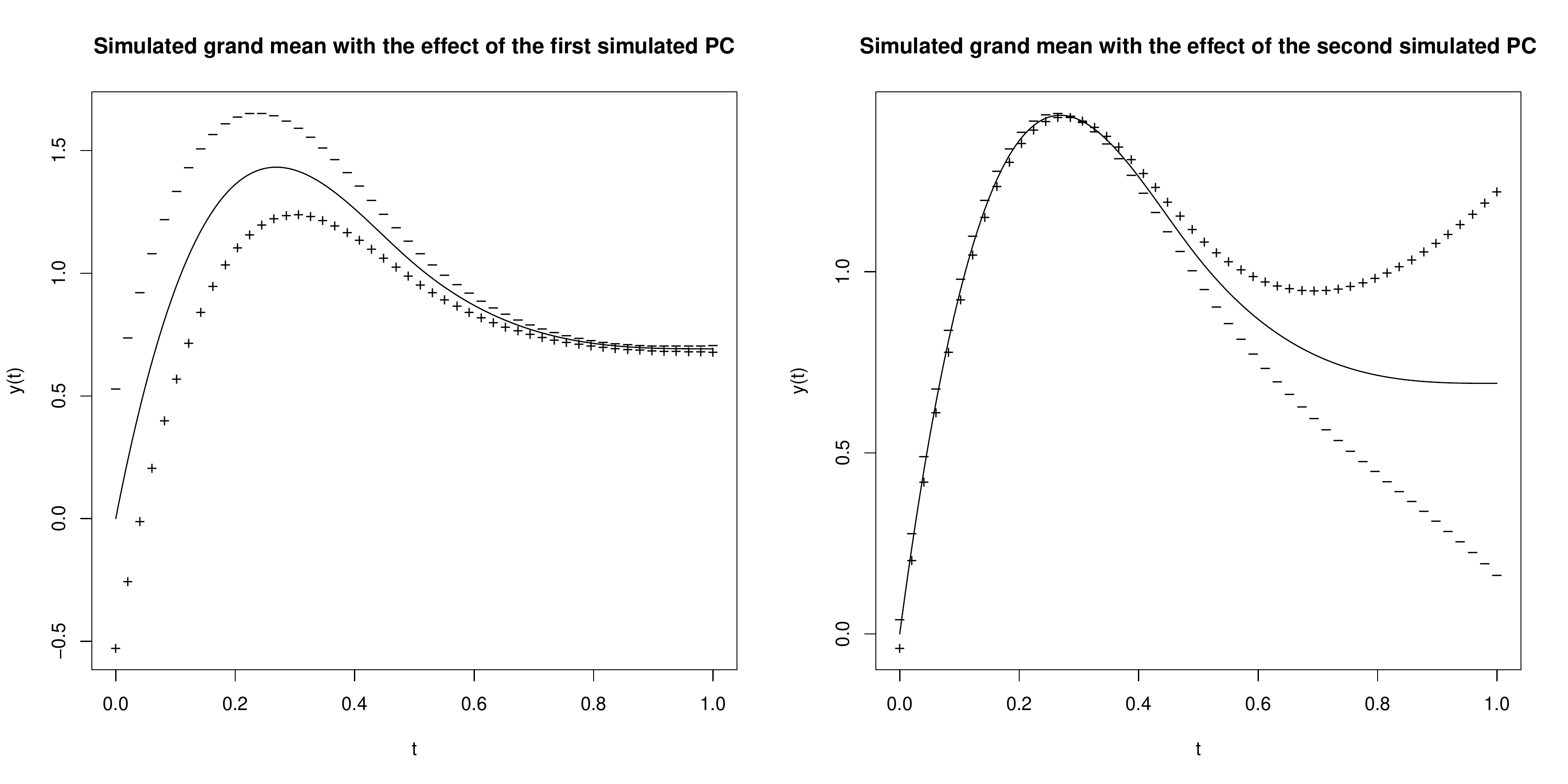}
\caption
{\label{fig:mlfpca simulation mean plus pcs}The effect of the two simulated principal
component functions on the simulated grand mean from our simulation study to compare the
Gaussian single- and
multi-level reduced-rank FPCA models. The first principal component function rotates the first half
of the time course, therefore affecting the height to which the functions peak, the level at which
the time course begins and, to a lesser extent, the exact time at which the peak occurs. Conversely
the second principal component function rotates the second half of the time course, thereby
controlling whether the curve levels off by the end of the time course, continues to decrease or
starts to increase.}
\end{figure}

The spline coefficients for the variable-level principal component functions,
$\bm{\theta}_{\alpha_1}$ and $\bm{\theta}_{\alpha_2}$ were chosen in order to produce the simulated
curves given in Figure \ref{fig:mlfpca simulation pcs}. Visualising the grand mean plus and minus
each principal component function is a typical way of helping to understand their effect
\citep{Ramsay2005}. These plots are given in Figure \ref{fig:mlfpca simulation mean plus pcs}. The
solid line is the grand mean. The points denoted by `$+$' are the function $\mu(t) + C\zeta_k(t)$
evaluated on a coarse grid of points where $C$ is some constant responsible for scaling the
principal component function and subjectively chosen in order to aid clarity of the visualisation.
Similarly, the points denoted by `$-$' are the same except for $\mu(t) - C\zeta_k(t)$. From these
plots it should be clear that the first principal component has the effect of rotating the first
half of the time course, which alters both the level at which the time course begins and the level
to which it peaks. To a lesser extent the exact time at which the peak occurs is also affected. On
the other hand, the second principal component function rotates the second half of the time course,
thereby controlling whether the curve levels off, continues to decrease or starts to increase after
the peak and the dip.

\begin{figure}[tbp]
\includegraphics[width=\textwidth]{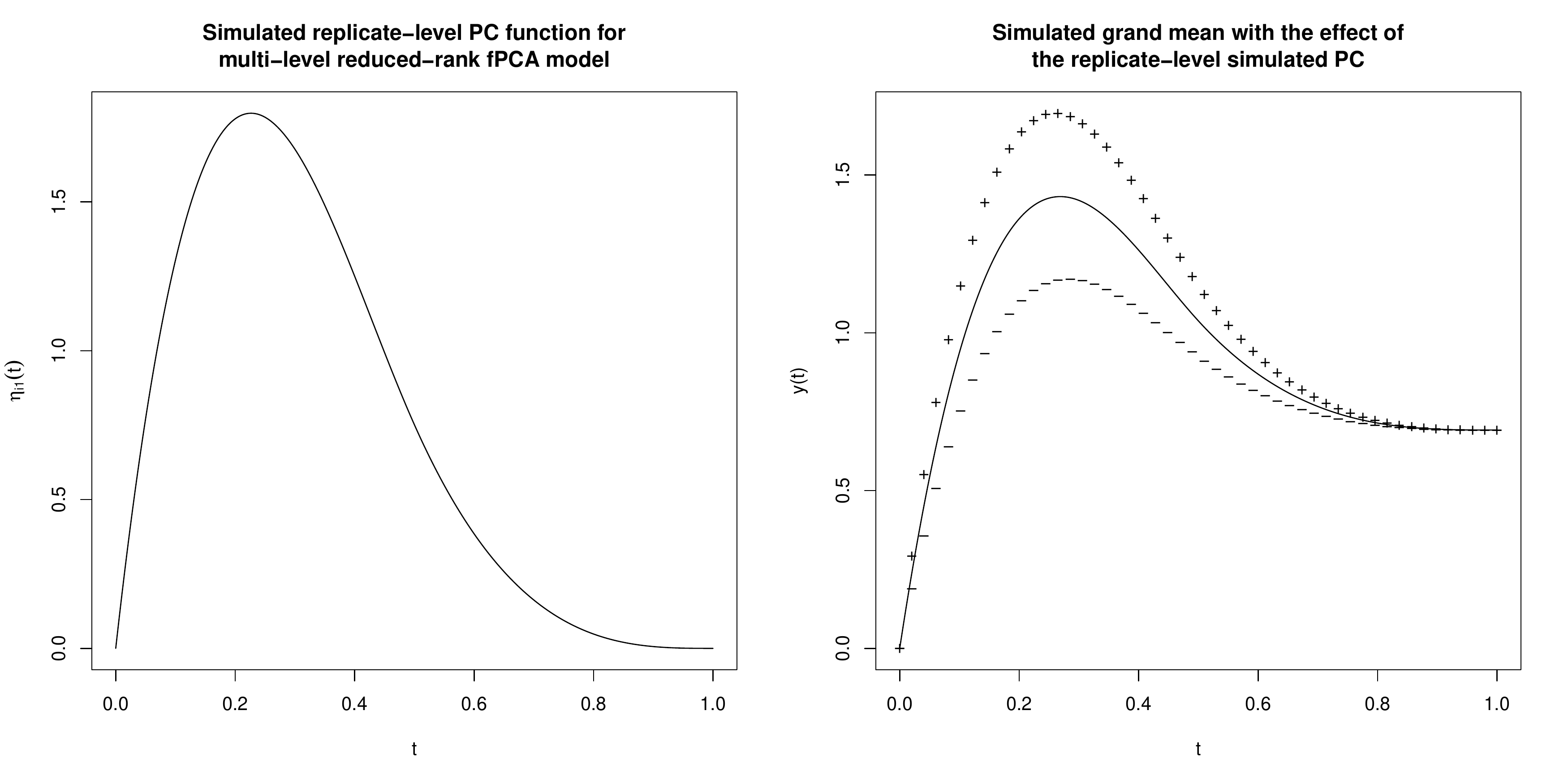}
\caption
{\label{fig:mlfpca simulation mean plus replicate pc}Profile of the simulated
replicate-level principal component function used in our simulation study to compare the Gaussian
single- and multi-level reduced-rank FPCA models, left, and its effect on the grand mean, right. The
principal component function has the effect of scaling the height to which the curve peaks.}
\end{figure}

For the replicate-level principal components, the same principal component function was fixed for
all variables. Note that this only serves to simplify the simulation scheme, and the multi-level
reduced-rank FPCA model will still estimate the between-replicate variation for each variable
independently. By choosing the spline coefficients $\bm{\theta}_{\beta_1}$ to produce the profile
given on the left hand side of Figure \ref{fig:mlfpca simulation mean plus replicate pc}, the effect
of the principal component function is to scale the height to which the curves peak, as shown on the
right hand side of the same Figure.

The remaining parameters to be determined are the variance components $\bm{D}_{\alpha}$,
$\bm{D}_{\beta_i}$ and $\sigma_{i}^{2}$ for all $i$. We set $\bm{D}_{\alpha} = \mbox{diag}(0.3,0.1)$
so that $75\%$ of the variance at the variable level is explained by the first principal component
function. The single element of $\bm{D}_{\beta_i}$ was set to $0.075$ for all $i$ so that the level
of between-replicate variance is less than that of the between-variable. Similarly, the noise was
set to $\sigma_{i}^{2} = 0.05$ for all $i$. As before, this only serves to simplify the simulation
scheme and separate estimates will be made for $\bm{D}_{\beta_i}$ and $\sigma_{i}^{2}$ for each
variable. The simulation scheme described here is flexible enough to produce a wide range of
believable curves as evidenced by the examples given in Supplementary Figure 5.

In order to compare the single- and multi-level reduced-rank FPCA models under a range of
experimental designs and to determine whether standard practice is appropriate, we generated
different data sets by varying the number of replicates as either $5$, $10$ or $20$. Roughly
speaking, with regards real data sets, these correspond to realistic, less frequent and unrealistic
numbers of replicates respectively. The number of time points per data set was fixed to $5$. We
focused on evaluating the ability to estimate the variable-level curves and for this reason we also
varied the number of variables between $100$, $1,000$ and $10,000$. Although broadly speaking
$10,000$ is the only realistic value of the three for `omics data sets (unless the variables have
been subject to some initial filtering procedure), we were interested in exploring the properties of
the multi-level model and determining how many variables are required to adequately assess the
between-variable variation.

We generated $1,000$ data sets under each condition. For each data set we fit both the single- and
multi-level reduced-rank FPCA models under the assumption of normality. For the purposes of this
study we chose not to consider the issue of correctly selecting the number of principal component
functions at the variable- and
replicate-levels and the number or location of the knots of the spline basis and simply input the
correct values to each algorithm. We compared each variable-level curve by discretising it on a fine
grid of points and calculating the mean squared error between it and the underlying true curve. We
then averaged this error across all variables and all data sets to give a single measure for each
model for each condition.

The complete results of the simulation study are presented in Supplementary Tables
3 and 4. In all scenarios the
multi-level model substantially improves upon the single-level model. For the multi-level model,
doubling the number of replicates roughly halves the estimation error. However, increasing the
number of variables has a much less pronounced effect, suggesting the multi-level model still
performs well when data sets have been pre-filtered. For the single-level model, doubling the number
of replicates similarly roughly halves the estimation error. As expected, increasing the number of
variables for this model has no effect on the estimation error as each variable is fit in isolation.
The most salient insight to glean from these results is in the comparison between the two tables.
Comparing first the case of $5$ replicates and $10,000$ variables for the two models, the
multi-level model offers an improvement in estimation error of approximately a factor of
ten. Considering the single-level model alone, in order to attain an equivalent improvement in
estimation error, the number of replicates needs to quadruple from $5$ to $20$. These results
suggest that simply using a more sophisticated model has the potential to dramatically reduce
experimental costs.

\subsection{Real data analysis}

We fit the skew-$t$-normal multi-level reduced-rank FPCA model to a genomics data set study the
genetic response to infection by BCG, the vaccine for tuberculosis, in $9$ human volunteers. We
determined that the MCEM algorithm had converged after $1050$ iterations by examining parameter
trace plots. On the right hand side of Figure \ref{fig:gaussian problem} we show the fit obtained
under the new model to the CCL20 transcript. As can be clearly seen, the mean curve is much more
reasonable, closely following the underlying observations.

\begin{figure}[htbp]
\includegraphics[width=\textwidth]{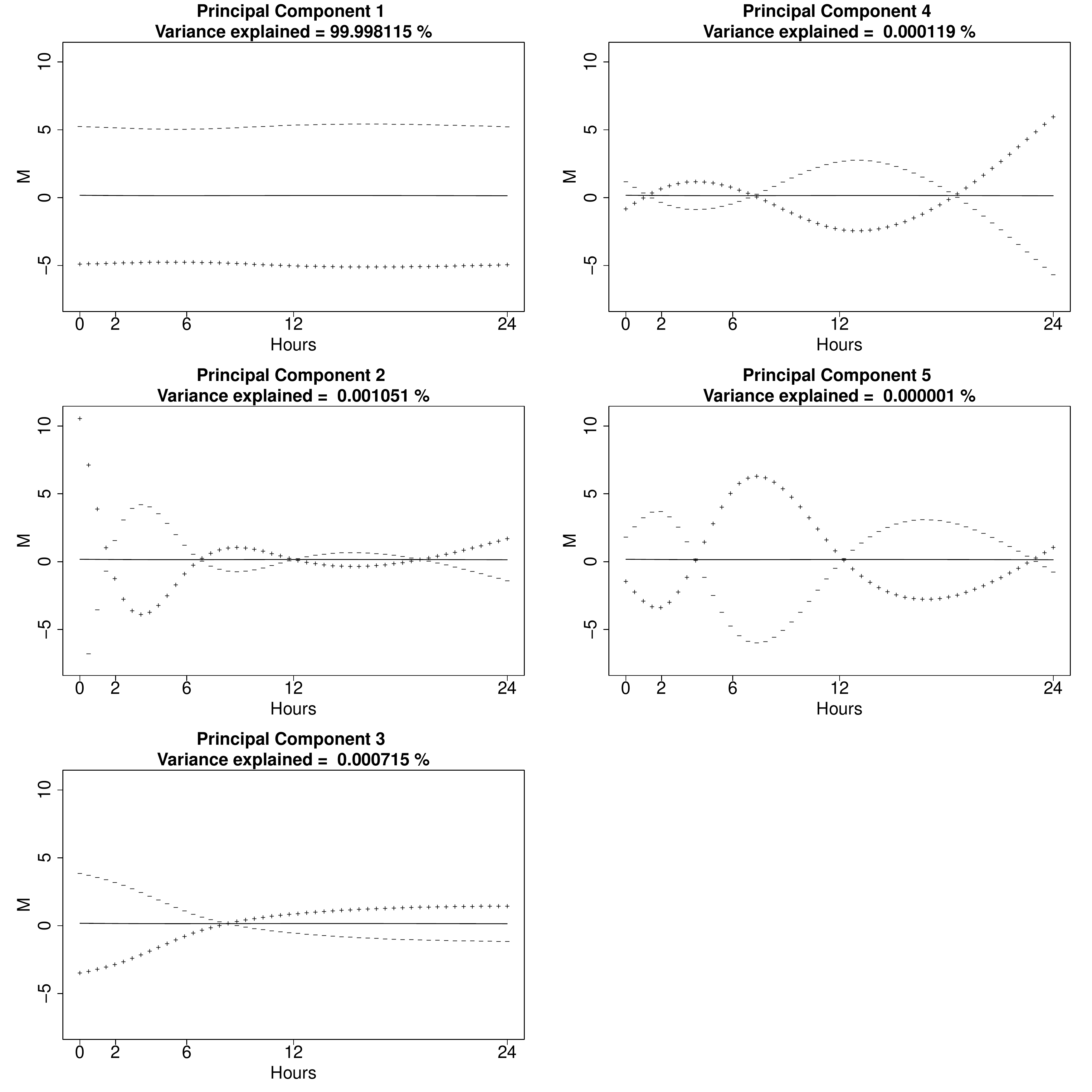}
\caption
{\label{fig:interpretable pcs}Variable-level principal component functions obtained from
fitting the skew-\textit{t}-normal multi-level reduced-rank fPCA model to our example real data
set. The solid line is the grand mean across all transcripts. The points `+' indicate the effect of
adding each principal component function multiplied by a constant $C$ to the grand mean, where $C$
is simply subjectively chosen to aid the visualisation. Similarly the points `-' indicate the effect
of subtracting each principal component function multiplied by $C$ from the grand mean. See
discussion in the main text.}
\end{figure}

The variable-level principal component functions are given in Figure \ref{fig:interpretable pcs}.
The first principal component -- which explains a huge proportion of the variance, $99.998\%$ -- is
responsible for vertically shifting a given transcript. In this respect, it is the most
uninteresting of the principal component functions as it does not control the shape of the curves.
We stress that it is not suprising that such a large proportion of the variance is explained by
vertical shifts given that very few of the tens of thousands of variables in an `omics data set will
actually exhibit significant changes over time. It is therefore still instructive to consider the
other principal component functions even if the proportion of variance they explain may lead to them
being overlooked in more traditional PCA application areas.

The second principal component function accounts for those transcripts which rapidly spike before
plateauing at some elevated level of expression or, conversely, are rapidly repressed. The CCL20
transcript is a example of this. The third principal component function describes those transcripts
which are induced or repressed more slowly, tailing off at around $7$ or $8$ hours. In Supplementary
Figure 6 we give an example of a transcript that exhibits this profile and was found to
have the highest positive loading on the third principal component function. More complex profiles
can be explained by a combination of these two components. The fourth principal component function
is less interpretable but appears to be mainly controlling for variation at the end of the time
course. The fifth principal component function is even less interpretable and should probably be
discarded when the data is fit for a second time under the reduced-rank model.

\section{Discussion}\label{sec:multivariate discussion}

To date, multi-level functional models such as the one we have presented here have yet to be applied
in the `omics fields that we focus on. In other domains, \cite{Zhou2010} independently developed a
similar multi-level reduced-rank FPCA model under the normality assumption. Their model differs from
ours in a few key respects. Firstly, the second-level prinicpal component loadings (in our model,
this would be the replicate-level) are not specific to the first-level. In other words, in our
context, their model would assume that the within-variable variation is identical for all variables.
As we motivated with Supplementary Figures 1 and 2, real
data does not support this assumption. Secondly, they allow for the replicate-level loadings to be
correlated within a given variable. In this respect their model is an extension of ours; however,
this is motivated by spatial dependencies in their case study that do not exist in our example data
sets. Thirdly, they use a penalised spline representation for the principal component functions. In
principle this should allow for a more data-driven approach to smoothing than our choice of natural
splines with the maximum number of knots. However, we note that their data set has many more `time'
points (in fact, the dependent variable is distance), than our case studies ($20-40$) and therefore
smoothing is more likely to be an issue if the function has been oversampled. Furthermore, they
select only three distinct smoothing parameters, one for the grand mean, one for all variable-level
principal component functions, and one for all replicate-level principal component functions.
Although it is clear to understand the computational burden that motivates such a restriction, it
seems to do away with the advantage that motivates penalised estimation in the first place, which is
to account for principal component functions of varying smoothness. Fourthly and finally, they fit a
single error variance for all variables. It is well-known that noise in microarray experiments is
transcript-dependent \citep{Tusher2001} and so such a noise model would be inadequate for our
purposes. Finally, the model is demonstrated on a data set with far fewer variables than our case
studies ($3$ comapred with on the order of $10,000$) which may explain why they did not experience
the same problems with assuming normality that we did.

% A different approach to multi-level FPCA can be found in \cite{Di2009}. In this approach, first the
% covariance functions of the design time points at both levels are estimated using the method of
% moments. These functions are then smoothed using thin plate smoothing in order to estimate the
% covariance surface over the entire range of the time course. Finally, an eigendecomposition is
% performed on the matrix formed by discretising the surfaces on a fine grid of points to obtain the
% principal component functions. It should be immediately obvious that the major drawback to such an
% approach is that it lacks the computational efficiency and stability of the reduced-rank model by
% estimating the full-rank covariance surface. Furthermore, the motivating case study contains over
% $3,000$ replicates and $960$ time points and it is unlikely the method of moments approach would
% yield good results with our far smaller sample sizes.

Although we have demonstrated that the MCEM algroithm for the skew-\textit{t}-normal model
yields biologically interpretable results on real data, the computational burden is severe. We
suggest several lines of attack for developing a practical algorithm that can fit the model in a
more reasonable time frame. Firstly, a different sampling scheme could be employed in the MCEM
algorithm, specifically importance sampling using a multivariate \textit{t} proposal with a small
degrees of freedom parameter. Moving away from the EM-algorithm, there may be the potential for an
(approximate) analytical solution. In particular, the work of \cite{Forchini2008} on deriving the
form of the density of the sum of a normal and a Student-\textit{t} distributed random variable may
provide some guidance. Deriving the density of the sum of a normal and a skew-\textit{t}-normal
random variable would be an important first step. However, the result of \cite{Forchini2008} relies
on truncating a Taylor series expansion and so careful consideration should be given as to the
accuracy of such an approximation. Ultimately, however, we believe that the most fruitful line of
further investigation is a fully Bayesian approach, taken by imposing prior distributions on the
model parameters. Parameter estimation could then be carried out using a modified version of our
Gibbs sampler. \cite{Jara2008} have already demonstrated that such a Bayesian approach works well
for single-level mixed-effects models with either skew-\textit{t} or skew-normal random-effects.

% AOS,AOAS: If there are supplements please fill:
%\begin{supplement}[id=suppA]
%  \sname{Supplementary Material}
%  \stitle{Title}
  %\slink[url]{http://lib.stat.cmu.edu/aoas/???/???}
%  \sdescription{In this file we provide additional technical details, including mathematical
%  derivations, and figures}
%\end{supplement}

\bibliographystyle{imsart-nameyear}

\bibliography{../References/references}

\begin{thebibliography}{24}
% BibTex style file: imsart-nameyear.bst, 2010-01-14
% Default style options (sort=1,type=nameyear).
% Used options (sort=1,type=nameyear).

\bibitem[\protect\citeauthoryear{Anderson}{1958}]{Anderson1958}
\begin{bbook}[author]
\bauthor{\bsnm{Anderson},~\bfnm{Theodore~W.}\binits{T.~W.}}
(\byear{1958}).
\btitle{Introduction to Multivariate Statistical Analysis}.
\bpublisher{Wiley}.
\end{bbook}
\endbibitem

\bibitem[\protect\citeauthoryear{Angelini, Canditiis and
  Pensky}{2009}]{Angelini2009}
\begin{barticle}[author]
\bauthor{\bsnm{Angelini},~\bfnm{Claudia}\binits{C.}},
  \bauthor{\bsnm{Canditiis},~\bfnm{Daniela~De}\binits{D.~D.}} \AND
  \bauthor{\bsnm{Pensky},~\bfnm{Marianna}\binits{M.}}
(\byear{2009}).
\btitle{Bayesian models for two-sample time-course microarray experiments}.
\bjournal{Computational Statistics \& Data Analysis}
\bvolume{53}
\bpages{1547 - 1565}.
\bnote{Statistical Genetics \& Statistical Genomics: Where Biology,
  Epistemology, Statistics, and Computation Collide}.
\end{barticle}
\endbibitem

\bibitem[\protect\citeauthoryear{Azzalini and Capitanio}{2003}]{Azzalini2003}
\begin{barticle}[author]
\bauthor{\bsnm{Azzalini},~\bfnm{A.}\binits{A.}} \AND
  \bauthor{\bsnm{Capitanio},~\bfnm{A.}\binits{A.}}
(\byear{2003}).
\btitle{Distributions generated by perturbation of symmetry with emphasis on a
  multivariate skew-t distribution}.
\bjournal{Journal of the Royal Statistical Society, Series B}
\bvolume{65}
\bpages{367-389}.
\end{barticle}
\endbibitem

\bibitem[\protect\citeauthoryear{Berk, Ebbels and Montana}{2011}]{Berk2011}
\begin{barticle}[author]
\bauthor{\bsnm{Berk},~\bfnm{M.}\binits{M.}},
  \bauthor{\bsnm{Ebbels},~\bfnm{T.}\binits{T.}} \AND
  \bauthor{\bsnm{Montana},~\bfnm{G.}\binits{G.}}
(\byear{2011}).
\btitle{A statistical framework for metabolic profiling using longitudinal
  data}.
\bjournal{Bioinformatics}
\bvolume{27}
\bpages{1979 - 1985}.
\end{barticle}
\endbibitem

\bibitem[\protect\citeauthoryear{Berk et~al.}{2010}]{Berk2010}
\begin{binproceedings}[author]
\bauthor{\bsnm{Berk},~\bfnm{M.}\binits{M.}},
  \bauthor{\bsnm{Montana},~\bfnm{G.}\binits{G.}},
  \bauthor{\bsnm{Levin},~\bfnm{M.}\binits{M.}} \AND
  \bauthor{\bsnm{Hemingway},~\bfnm{C.}\binits{C.}}
(\byear{2010}).
\btitle{Longitudinal analysis of gene expression profiles using functional
  mixed-effects models}.
In \bbooktitle{Studies in Theoretical and Applied Statistics}.
\end{binproceedings}
\endbibitem

\bibitem[\protect\citeauthoryear{Di et~al.}{2009}]{Di2009}
\begin{barticle}[author]
\bauthor{\bsnm{Di},~\bfnm{C.}\binits{C.}},
  \bauthor{\bsnm{Crainiceanu},~\bfnm{C.~M.}\binits{C.~M.}},
  \bauthor{\bsnm{Kuechenhoff},~\bfnm{H.}\binits{H.}} \AND
  \bauthor{\bsnm{Peters},~\bfnm{A.}\binits{A.}}
(\byear{2009}).
\btitle{Multilevel functional principal component analysis}.
\bjournal{Annals of Applied Statistics}
\bvolume{3}
\bpages{458 - 488}.
\end{barticle}
\endbibitem

\bibitem[\protect\citeauthoryear{Forchini}{2008}]{Forchini2008}
\begin{barticle}[author]
\bauthor{\bsnm{Forchini},~\bfnm{G.}\binits{G.}}
(\byear{2008}).
\btitle{The distribution of the sum of a normal and a t random variable with
  arbitrary degrees of freedom}.
\bjournal{{METRON - International Journal of Statistics}}
\bvolume{2}
\bpages{205-208}.
\end{barticle}
\endbibitem

\bibitem[\protect\citeauthoryear{G\'{o}mez, Venegas and
  Bolfarine}{2007}]{G'omez2007}
\begin{barticle}[author]
\bauthor{\bsnm{G\'{o}mez},~\bfnm{Hector~W.}\binits{H.~W.}},
  \bauthor{\bsnm{Venegas},~\bfnm{Osvaldo}\binits{O.}} \AND
  \bauthor{\bsnm{Bolfarine},~\bfnm{Heleno}\binits{H.}}
(\byear{2007}).
\btitle{Skew-symmetric distributions generated by the distribution function of
  the normal distribution}.
\bjournal{Environmetrics}
\bvolume{18}
\bpages{395--407}.
\end{barticle}
\endbibitem

\bibitem[\protect\citeauthoryear{Ho and Lin}{2010}]{Ho2010}
\begin{barticle}[author]
\bauthor{\bsnm{Ho},~\bfnm{Hsiu~J.}\binits{H.~J.}} \AND
  \bauthor{\bsnm{Lin},~\bfnm{Tsung-I.}\binits{T.-I.}}
(\byear{2010}).
\btitle{Robust linear mixed models using the skew t distribution with
  application to schizophrenia data}.
\bjournal{Biometrical Journal}
\bvolume{52}
\bpages{449 - 469}.
\end{barticle}
\endbibitem

\bibitem[\protect\citeauthoryear{James, Hastie and Sugar}{2000}]{James2000}
\begin{barticle}[author]
\bauthor{\bsnm{James},~\bfnm{GM}\binits{G.}},
  \bauthor{\bsnm{Hastie},~\bfnm{TJ}\binits{T.}} \AND
  \bauthor{\bsnm{Sugar},~\bfnm{CA}\binits{C.}}
(\byear{2000}).
\btitle{{Principal component models for sparse functional data}}.
\bjournal{Biometrika}
\bvolume{87}
\bpages{587-602}.
\end{barticle}
\endbibitem

\bibitem[\protect\citeauthoryear{Jara, Quintana and Martin}{2008}]{Jara2008}
\begin{barticle}[author]
\bauthor{\bsnm{Jara},~\bfnm{Alejandro}\binits{A.}},
  \bauthor{\bsnm{Quintana},~\bfnm{Fernando}\binits{F.}} \AND
  \bauthor{\bsnm{Martin},~\bfnm{Ernesto~San}\binits{E.~S.}}
(\byear{2008}).
\btitle{{Linear mixed models with skew-elliptical distributions: A Bayesian
  approach}}.
\bjournal{Computational Statistics \& Data Analysis}
\bvolume{52}
\bpages{5033 - 5045}.
\end{barticle}
\endbibitem

\bibitem[\protect\citeauthoryear{Liu and Yang}{2009}]{Liu2009}
\begin{barticle}[author]
\bauthor{\bsnm{Liu},~\bfnm{Xueli}\binits{X.}} \AND
  \bauthor{\bsnm{Yang},~\bfnm{Mark C.~K.}\binits{M.~C.~K.}}
(\byear{2009}).
\btitle{{Identifying temporally differentially expressed genes through
  functional principal components analysis}}.
\bjournal{Biostat}
\bpages{kxp022}.
\end{barticle}
\endbibitem

\bibitem[\protect\citeauthoryear{Ma et~al.}{2006}]{Ma2006}
\begin{barticle}[author]
\bauthor{\bsnm{Ma},~\bfnm{Ping}\binits{P.}},
  \bauthor{\bsnm{Castillo-Davis},~\bfnm{Cristian~I}\binits{C.~I.}},
  \bauthor{\bsnm{Zhong},~\bfnm{Wenxuan}\binits{W.}} \AND
  \bauthor{\bsnm{Liu},~\bfnm{Jun~S}\binits{J.~S.}}
(\byear{2006}).
\btitle{A data-driven clustering method for time course gene expression data.}
\bjournal{Nucleic Acids Res}
\bvolume{34}
\bpages{1261--1269}.
\end{barticle}
\endbibitem

\bibitem[\protect\citeauthoryear{Montana, Berk and Ebbels}{2011}]{Montana2011}
\begin{binbook}[author]
\bauthor{\bsnm{Montana},~\bfnm{G.}\binits{G.}},
  \bauthor{\bsnm{Berk},~\bfnm{M.}\binits{M.}} \AND
  \bauthor{\bsnm{Ebbels},~\bfnm{T.}\binits{T.}}
(\byear{2011}).
\btitle{Software Tools and Algorithms for Biological Systems}
\bchapter{Modelling short time series in metabolomics: a functional data
  analysis approach}
\bpages{307-316}.
\bpublisher{Springer}, \baddress{New York}.
\end{binbook}
\endbibitem

\bibitem[\protect\citeauthoryear{Nelder and Mead}{1965}]{Nelder1965}
\begin{barticle}[author]
\bauthor{\bsnm{Nelder},~\bfnm{J.~A.}\binits{J.~A.}} \AND
  \bauthor{\bsnm{Mead},~\bfnm{R.}\binits{R.}}
(\byear{1965}).
\btitle{{A Simplex Method for Function Minimization}}.
\bjournal{The Computer Journal}
\bvolume{7}
\bpages{308-313}.
\end{barticle}
\endbibitem

\bibitem[\protect\citeauthoryear{Peng and Paul}{2009}]{Peng2009}
\begin{barticle}[author]
\bauthor{\bsnm{Peng},~\bfnm{Jie}\binits{J.}} \AND
  \bauthor{\bsnm{Paul},~\bfnm{Debashis}\binits{D.}}
(\byear{2009}).
\btitle{A Geometric Approach to Maximum Likelihood Estimation of the Functional
  Principal Components From Sparse Longitudinal Data}.
\bjournal{Journal of Computational and Graphical Statistics}
\bvolume{18}
\bpages{995-1015}.
\end{barticle}
\endbibitem

\bibitem[\protect\citeauthoryear{Ramsay and Silverman}{2005}]{Ramsay2005}
\begin{bbook}[author]
\bauthor{\bsnm{Ramsay},~\bfnm{Jim}\binits{J.}} \AND
  \bauthor{\bsnm{Silverman},~\bfnm{B.~W.}\binits{B.~W.}}
(\byear{2005}).
\btitle{Functional Data Analysis}, \bedition{2} ed.
\bpublisher{Springer}, \baddress{New York}.
\end{bbook}
\endbibitem

\bibitem[\protect\citeauthoryear{Storey et~al.}{2005}]{Storey2005}
\begin{barticle}[author]
\bauthor{\bsnm{Storey},~\bfnm{John~D}\binits{J.~D.}},
  \bauthor{\bsnm{Xiao},~\bfnm{Wenzhong}\binits{W.}},
  \bauthor{\bsnm{Leek},~\bfnm{Jeffrey~T}\binits{J.~T.}},
  \bauthor{\bsnm{Tompkins},~\bfnm{Ronald~G}\binits{R.~G.}} \AND
  \bauthor{\bsnm{Davis},~\bfnm{Ronald~W}\binits{R.~W.}}
(\byear{2005}).
\btitle{Significance analysis of time course microarray experiments.}
\bjournal{Proc Natl Acad Sci U S A}
\bvolume{102}
\bpages{12837--12842}.
\end{barticle}
\endbibitem

\bibitem[\protect\citeauthoryear{Tai and Speed}{2009}]{Tai2009}
\begin{barticle}[author]
\bauthor{\bsnm{Tai},~\bfnm{Yu~Chuan}\binits{Y.~C.}} \AND
  \bauthor{\bsnm{Speed},~\bfnm{Terence~P.}\binits{T.~P.}}
(\byear{2009}).
\btitle{On Gene Ranking Using Replicated Microarray Time Course Data}.
\bjournal{Biometrics}
\bvolume{65}
\bpages{40-51}.
\end{barticle}
\endbibitem

\bibitem[\protect\citeauthoryear{Tusher, Tibshirani and Chu}{2001}]{Tusher2001}
\begin{barticle}[author]
\bauthor{\bsnm{Tusher},~\bfnm{Virginia~Goss}\binits{V.~G.}},
  \bauthor{\bsnm{Tibshirani},~\bfnm{Robert}\binits{R.}} \AND
  \bauthor{\bsnm{Chu},~\bfnm{Gilbert}\binits{G.}}
(\byear{2001}).
\btitle{{Significance analysis of microarrays applied to the ionizing radiation
  response}}.
\bjournal{Proceedings of the National Academy of Sciences of the United States
  of America}
\bvolume{98}
\bpages{5116-5121}.
\end{barticle}
\endbibitem

\bibitem[\protect\citeauthoryear{Wei and Tanner}{1990}]{Wei1990}
\begin{barticle}[author]
\bauthor{\bsnm{Wei},~\bfnm{Greg C.~G.}\binits{G.~C.~G.}} \AND
  \bauthor{\bsnm{Tanner},~\bfnm{Martin~A.}\binits{M.~A.}}
(\byear{1990}).
\btitle{{A Monte Carlo Implementation of the EM Algorithm and the Poor Man's
  Data Augmentation Algorithms}}.
\bjournal{Journal of the American Statistical Association}
\bvolume{85}
\bpages{pp. 699-704}.
\end{barticle}
\endbibitem

\bibitem[\protect\citeauthoryear{Yao, M{\"u}ller and Wang}{2005}]{Yao2005}
\begin{barticle}[author]
\bauthor{\bsnm{Yao},~\bfnm{Fang}\binits{F.}},
  \bauthor{\bsnm{M{\"u}ller},~\bfnm{Hans-Georg}\binits{H.-G.}} \AND
  \bauthor{\bsnm{Wang},~\bfnm{Jane-Ling}\binits{J.-L.}}
(\byear{2005}).
\btitle{Functional Data Analysis for Sparse Longitudinal Data}.
\bjournal{Journal of the American Statistical Association}
\bvolume{100}
\bpages{577-590}.
\end{barticle}
\endbibitem

\bibitem[\protect\citeauthoryear{Zhou, Huang and Carroll}{2008}]{Zhou2008}
\begin{barticle}[author]
\bauthor{\bsnm{Zhou},~\bfnm{Lan}\binits{L.}},
  \bauthor{\bsnm{Huang},~\bfnm{Jianhua~Z.}\binits{J.~Z.}} \AND
  \bauthor{\bsnm{Carroll},~\bfnm{Raymond~J.}\binits{R.~J.}}
(\byear{2008}).
\btitle{{Joint modelling of paired sparse functional data using principal
  components}}.
\bjournal{Biometrika}
\bvolume{95}
\bpages{601-619}.
\end{barticle}
\endbibitem

\bibitem[\protect\citeauthoryear{Zhou et~al.}{2010}]{Zhou2010}
\begin{barticle}[author]
\bauthor{\bsnm{Zhou},~\bfnm{Lan}\binits{L.}},
  \bauthor{\bsnm{Huang},~\bfnm{Jianhua~Z.}\binits{J.~Z.}},
  \bauthor{\bsnm{Martinez},~\bfnm{Josue~G.}\binits{J.~G.}},
  \bauthor{\bsnm{Maity},~\bfnm{Arnab}\binits{A.}},
  \bauthor{\bsnm{Baladandayuthapani},~\bfnm{Veerabhadran}\binits{V.}} \AND
  \bauthor{\bsnm{Carroll},~\bfnm{Raymond~J.}\binits{R.~J.}}
(\byear{2010}).
\btitle{Reduced Rank Mixed Effects Models for Spatially Correlated Hierarchical
  Functional Data}.
\bjournal{Journal of the American Statistical Association}
\bvolume{105}
\bpages{390-400}.
\end{barticle}
\endbibitem

\end{thebibliography}


\begin{thebibliography}{4}
\providecommand{\natexlab}[1]{#1}
\providecommand{\url}[1]{\texttt{#1}}
\expandafter\ifx\csname urlstyle\endcsname\relax
  \providecommand{\doi}[1]{doi: #1}\else
  \providecommand{\doi}{doi: \begingroup \urlstyle{rm}\Url}\fi

\bibitem[Anderson(1958)]{Anderson1958}
T.~W. Anderson.
\newblock \emph{Introduction to Multivariate Statistical Analysis}.
\newblock Wiley, 1958.

\bibitem[James et~al.(2000)James, Hastie, and Sugar]{James2000}
G.~James, T.~Hastie, and C.~Sugar.
\newblock {Principal component models for sparse functional data}.
\newblock \emph{Biometrika}, 87\penalty0 (3):\penalty0 587--602, 2000.

\bibitem[Peng and Paul(2009)]{Peng2009}
J.~Peng and D.~Paul.
\newblock A geometric approach to maximum likelihood estimation of the
  functional principal components from sparse longitudinal data.
\newblock \emph{Journal of Computational and Graphical Statistics}, 18\penalty0
  (4):\penalty0 995--1015, 2009.

\bibitem[Zhou et~al.(2010)Zhou, Huang, Martinez, Maity, Baladandayuthapani, and
  Carroll]{Zhou2010}
L.~Zhou, J.~Z. Huang, J.~G. Martinez, A.~Maity, V.~Baladandayuthapani, and
  R.~J. Carroll.
\newblock Reduced rank mixed effects models for spatially correlated
  hierarchical functional data.
\newblock \emph{Journal of the American Statistical Association}, 105\penalty0
  (489):\penalty0 390--400, 2010.

\end{thebibliography}

\end{document}

\title{Supplementary Material for ``A Skew-$t$-Normal Multi-Level Reduced-Rank Functional PCA Model
with Applications to Replicated `Omics Time Series Data Sets''}
\date{}
\maketitle

\section{The Gaussian multi-level reduced-rank FPCA model}

For the Gaussian case we take the following distributional assumptions:
\begin{align}
\bm{\alpha}_{i} \stackrel{i.i.d.}{\sim} MVN(\bm{0},\bm{D}_{\alpha}) \quad \bm{\beta}_{ij}
\stackrel{i.i.d.}{\sim} MVN(\bm{0}, \bm{D}_{\beta_i}) \quad \bm{\epsilon}_{ij}
\stackrel{i.i.d.}{\sim} MVN(\bm{0},\sigma_{i}^{2}\bm{I}_{N_{ij} \times N_{ij}})
\label{eqn:gaussian mlfpca assumptions}
\end{align}
where the $K \times K$ and $L_i \times L_i$ matrices $\bm{D}_{\alpha}$ and $\bm{D}_{\beta_{i}}$ are
both assumed to be diagonal otherwise they would be confounded with $\bm{\Theta}_{\alpha}$ and
$\bm{\Theta}_{\beta_{i}}$. Furthermore, we
assume that $\bm{\alpha}_{i}$, $\bm{\beta}_{ij}$ and $\bm{\epsilon}_{ij}$ are all independent of
each other for all $i$ and $j$. To enforce the orthogonality constraint on the functions
$\zeta_k(t)$ and $\eta_{il}(t)$, in addition to transforming the spline basis, we also impose that
$\bm{\Theta}_{\alpha}^{T}\bm{\Theta}_{\alpha} = \bm{I}_{K \times K}$ and
$\bm{\Theta}_{\beta_{i}}^{T}\bm{\Theta}_{\beta_{i}} = \bm{I}_{L_{i} \times L_{i}}$.

Under the distributional assumptions given above, $\bm{y}_{i}$ is marginally distributed as
\begin{align*}
\bm{y}_{i} &\sim MVN(\bm{B}_{i}\bm{\theta}_{\mu},\bm{V}_{i})
\end{align*}
where
\begin{align}
\label{eqn:gaussian mlfpca Vi}
\bm{V}_{i} &= \bm{B}_{i}\bm{\Theta}_{\alpha}\bm{D}_{\alpha}\bm{\Theta}_{\alpha}^{T}
\bm{B}_{i}^{T} + \widetilde{\bm{B}_{i}}\widetilde{\bm{\Theta}_{\beta_i}}\widetilde{\bm{D}_{\beta_i}}
\widetilde{\bm{\Theta}_{\beta_i}^{T}}\widetilde{\bm{B}_{i}^{T}} + \sigma_{i}^{2}
\bm{I}_{N_i \times N_i} \\
\nonumber \widetilde{\bm{D}_{\beta_i}} &= \mbox{diag}(\bm{D}_{\beta_i},\cdots,
\bm{D}_{\beta_i})
\end{align}
As in \cite{James2000}, the model parameters
\begin{equation*}
\bm{\psi}_i = \left\{ \bm{\theta}_{\mu}, \bm{\Theta}_{\alpha}, \bm{\Theta}_{\beta_{i}},
\bm{D}_{\alpha}, \bm{D}_{\beta_{i}} \sigma_{i}^{2} \right\} \quad i=1,\cdots,M
\end{equation*}
where $M$ is the total
number of variables in the data set, can be estimated by treating the principal component loadings
as missing data and employing the EM algorithm. Under the distributional assumptions given above,
the complete data log-likelihood is
\begin{align}
\nonumber \sum_{i=1}^{M} \mathcal{L}(\bm{\psi}_i|\bm{y}_{i}) =& 
\sum_{i=1}^{M} \left[ \log f(\bm{y}_{i}|\bm{\alpha}_{i},\bm{\beta}_{i},\bm{\psi}_{i}) +
\log f(\bm{\alpha}_{i}|\bm{\psi}_{i}) + \sum_{j=1}^{n_{i}} \log f(\bm{\beta}_{ij}|\bm{\psi}_{i})
\right] \\
\nonumber = C - \frac{1}{2} &\sum_{i=1}^{M}\Bigg[ N_i \log \sigma_{i}^{2} +
\sigma_{i}^{-2}||\bm{y}_{i} -
\bm{B}_{i}\bm{\theta}_{\mu} -\bm{B}_{i}\bm{\Theta}_{\alpha}\bm{\alpha}_{i} - \widetilde{\bm{B}_{i}}
\widetilde{\bm{\Theta}_{\beta_i}}\bm{\beta}_{i}||^2 + \\
& \log |\bm{D}_{\alpha}| + \bm{\alpha}_{i}^{T}\bm{D}_{\alpha}^{-1}\bm{\alpha}_{i} +
\sum_{j=1}^{n_{i}}\left[\log|\bm{D}_{\beta_i}| +
\bm{\beta}_{ij}^{T}\bm{D}_{\beta_{i}}^{-1}\bm{\beta}_{ij}\right]\Bigg]
\label{eqn:multi-level gaussian likelihood}
\end{align}
where $C$ is an additive constant. We will now proceed to derive the maximum likelihood estimators
of the model parameters, followed by the required conditional expectations for the EM algorithm.

\subsection{MLE of \texorpdfstring{$\bm{\theta}_{\mu}$}{\theta_{\mu}}}
%\subsection{MLE of $\theta_{\mu}$}

To derive the maximum likelihood estimator of $\bm{\theta}_{\mu}$ first ignore all irrelevant terms
in (\ref{eqn:multi-level gaussian likelihood}) and for succinctness write $\bm{y}_{i} -
\bm{B}_{i}\bm{\theta}_{\mu} -\bm{B}_{i}\bm{\Theta}_{\alpha}\bm{\alpha}_{i} - \widetilde{\bm{B}_{i}}
\widetilde{\bm{\Theta}_{\beta_i}}\bm{\beta}_{i} = \bm{y}_{i}^{*} - \bm{B}_{i}\bm{\theta}_{\mu}$.
Then the partial derivative with respect to $\bm{\theta}_{\mu}$ yields
\begin{align*}
\frac{ \partial \displaystyle -\frac{1}{2}\sum_{i=1}^{M}\sigma_{i}^{-2}(\bm{y}_{i}^{*} -
\bm{B}_{i}\bm{\theta}_{\mu})^{T}(\bm{y}_{i}^{*} - \bm{B}_{i}\bm{\theta}_{\mu})}{\partial
\bm{\theta}_{\mu}} &=
\sum_{i=1}^{M}\sigma_{i}^{-2}\left[-2\bm{B}_{i}^{T}\bm{y}_{i}^{*} + 2\bm{B}_{i}^{T}\bm{B}_{i}
\bm{\theta}_{\mu}\right]
\end{align*}
Equating to zero and solving for $\bm{\theta}_{\mu}$ gives
\begin{align}
\nonumber \sum_{i=1}^{M} (\bm{B}_{i}^{T}\bm{B}_{i})\bm{\theta}_{\mu} &= \sum_{i=1}^{M}
\bm{B}_{i}^{T}\bm{y}_{i}^{*} \\
\nonumber \widehat{\bm{\theta}_{\mu}} &= (\sum_{i=1}^{M}\bm{B}_{i}^{T}\bm{B}_{i})^{-1}
\sum_{i=1}^{M}\bm{B}_{i}^{T}\bm{y}_{i}^{*} \\
\widehat{\bm{\theta}_{\mu}} &= (\sum_{i=1}^{M}\bm{B}_{i}^{T}\bm{B}_{i})^{-1}
\sum_{i=1}^{M}\bm{B}_{i}^{T}\left[\bm{y}_{i} -\bm{B}_{i}\bm{\Theta}_{\alpha}\bm{\alpha}_{i} -
\widetilde{\bm{B}_{i}}\widetilde{\bm{\Theta}_{\beta_i}}\bm{\beta}_{i}\right]
\label{eqn:gaussian mlfpca thetamu mle}
\end{align}

\subsection{MLE of \texorpdfstring{$\bm{\theta}_{\alpha_k}$}{\theta_{\alpha_k}}}
%\subsection{MLE of $\theta_{\alpha_k}$}

For the maximum likelihood estimator of $\bm{\theta}_{\alpha_k}$, which recall is the $k$-th column
of $\bm{\Theta}_{\alpha}$, proceed in exactly the same way as for $\bm{\theta}_{\mu}$. For clarity
first redefine $\bm{y}_{i}^{*} = \bm{y}_{i} - \bm{B}_{i}\bm{\theta}_{\mu} - \bm{B}_{i}
\sum_{k' \ne k}\left[\bm{\theta}_{\alpha_{k'}}\alpha_{ik'}\right] - \widetilde{
\bm{\Theta}_{\beta_i}}\bm{\beta}_{i}$. Then, taking the partial derivative of the relevant terms of
(\ref{eqn:multi-level gaussian likelihood}) with respect to $\bm{\theta}_{\alpha_k}$ gives
\begin{align*}
\frac{ \partial \displaystyle -\frac{1}{2}\sum_{i=1}^{M}\sigma_{i}^{-2}(\bm{y}_{i}^{*} -
 \bm{B}_{i}\bm{\theta}_{\alpha_k}\alpha_{ik})^{T}(\bm{y}_{i}^{*} -
 \bm{B}_{i}\bm{\theta}_{\alpha_k}\alpha_{ik})}{\partial
 \bm{\theta}_{\alpha_k}} &= \\
 \sum_{i=1}^{M}\sigma_{i}^{-2} \bigg[ & 
 -2\alpha_{ik}\bm{B}_{i}^{T}\bm{y}_{i}^{*} + 2\alpha_{ik}^{2}\bm{B}_{i}^{T}\bm{B}_{i}
 \bm{\theta}_{\mu}\bigg]
\end{align*}
Equating to zero and solving for $\bm{\theta}_{\alpha_k}$ gives
\begin{align*}
\nonumber \sum_{i=1}^{M} (\alpha_{ik}^{2}\bm{B}_{i}^{T}\bm{B}_{i})\bm{\theta}_{\alpha_k} &=
\sum_{i=1}^{M}\alpha_{ik}\bm{B}_{i}^{T}\bm{y}_{i}^{*} \\
\nonumber \widehat{\bm{\theta}_{\alpha_k}} &= (\sum_{i=1}^{M}\alpha_{ik}^{2}\bm{B}_{i}^{T}
\bm{B}_{i})^{-1}
\sum_{i=1}^{M}\alpha_{ik}\bm{B}_{i}^{T}\bm{y}_{i}^{*}
\end{align*}
and so
\begin{align}
\widehat{\bm{\theta}_{\alpha_k}} &= (\sum_{i=1}^{M}\alpha_{ik}^{2}\bm{B}_{i}^{T}\bm{B}_{i})^{-1}
\sum_{i=1}^{M}\alpha_{ik}\bm{B}_{i}^{T}\left[\bm{y}_{i} - \bm{B}_{i}\bm{\theta}_{\mu}
- \bm{B}_{i}\sum_{k' \ne k}\left[\bm{\theta}_{\alpha_k}\alpha_{ik'}\right] -
\widetilde{\bm{B}_{i}}\widetilde{\bm{\Theta}_{\beta_i}}\bm{\beta}_{i}\right]
\label{eqn:gaussian mlfpca thetaalpha mle}
\end{align}

\subsection{MLE of \texorpdfstring{$\bm{\theta}_{\beta_{il}}$}{\theta_{\beta_{il}}}}
%\subsection{MLE of $\theta_{\beta_{il}}$}

The derivation for the maximum likelihood estimator of $\bm{\theta}_{\beta_{il}}$ is similar to that
for $\bm{\theta}_{\alpha_k}$ except this time we only need to consider observations on variable $i$.
Hence we define $\bm{y}_{ij}^{*} = \bm{y}_{ij} - \bm{B}_{ij}\bm{\theta}_{\mu} - \bm{B}_{ij}
\bm{\Theta}_{\alpha_{i}} - \bm{B}_{ij}\sum_{l' \ne l}\left[\bm{\theta}_{\beta_{il'}}\beta_{ijl'}
\right]$ and take the partial derivative of the relevant terms in
(\ref{eqn:multi-level gaussian likelihood}) with respect to $\bm{\theta}_{\beta_{il}}$ which yields
\begin{align*}
\frac{ \partial \displaystyle -\frac{1}{2}\sum_{j=1}^{n_{i}}\sigma_{i}^{-2}(\bm{y}_{ij}^{*} -
\bm{B}_{ij}\bm{\theta}_{\beta_{il}}\beta_{ijl})^{T}(\bm{y}_{i}^{*} -
\bm{B}_{i}\bm{\theta}_{\beta_{il}}\beta_{ijl})}{\partial
\bm{\theta}_{\beta_{il}}} &= \\
\sum_{j=1}^{n_i}\sigma_{i}^{-2} \bigg[ 
-2\beta_{ijl}\bm{B}_{ij}^{T}\bm{y}_{ij}^{*} & + 2\beta_{ijl}^{2}\bm{B}_{ij}^{T}\bm{B}_{ij}
\bm{\theta}_{\beta_{il}}\bigg]
\end{align*}
Equating to zero and solving for $\bm{\theta}_{\beta_{il}}$ gives
\begin{align*}
\nonumber \sum_{j=1}^{n_i} (\beta_{ijl}^{2}\bm{B}_{ij}^{T}\bm{B}_{ij})\bm{\theta}_{\beta_{il}} &=
\sum_{j=1}^{n_i}\beta_{ijl}\bm{B}_{ij}^{T}\bm{y}_{ij}^{*} \\
\nonumber \widehat{\bm{\theta}_{\beta_{il}}} &= (\sum_{j=1}^{n_i}\beta_{ijl}^{2}\bm{B}_{ij}^{T}
\bm{B}_{ij})^{-1}
\sum_{j=1}^{n_i}\beta_{ijl}\bm{B}_{ij}^{T}\bm{y}_{ij}^{*}
\end{align*}
and so
\begin{align}
\nonumber \widehat{\bm{\theta}_{\beta_{il}}} &= \\
 ( \sum_{j=1}^{n_i} & \beta_{ijl}^{2}
\bm{B}_{ij}^{T}\bm{B}_{ij})^{-1}
\sum_{j=1}^{n_i}\beta_{ijl}\bm{B}_{ij}^{T}\left[\bm{y}_{ij} - \bm{B}_{ij}\bm{\theta}_{\mu}
- \bm{B}_{ij}\bm{\Theta}_{\alpha}\bm{\alpha}_{i} - \bm{B}_{ij}\sum_{l' \ne l}\left[
\bm{\theta}_{\beta_il'}\beta_{ijl'}\right]\right]
\label{eqn:gaussian mlfpca thetabeta mle}
\end{align}

\subsection{MLE of \texorpdfstring{$\bm{D}_{\alpha}$}{D_{\alpha}}}
%\subsection{MLE of $D_{\alpha}$}

As $\bm{D}_{\alpha}$ is diagonal we consider estimating each diagonal element separately. The
relevant terms in (\ref{eqn:multi-level gaussian likelihood}) are $-\frac{1}{2}\sum_{i=1}^{M}
\left[\log |\bm{D}_{\alpha}| + \bm{\alpha}_{i}^{T}\bm{D}_{\alpha}^{-1}\bm{\alpha}_{i}\right]$. As
$\bm{D}_{\alpha}$ is diagonal we can write
\begin{align*}
-\frac{1}{2}\sum_{i=1}^{M}
\left[\log |\bm{D}_{\alpha}| + \bm{\alpha}_{i}^{T}\bm{D}_{\alpha}^{-1}\bm{\alpha}_{i}\right] &=
-\frac{1}{2}\sum_{i=1}^{M}
\left[\log \prod_{k'=1}^{K}\left[
\bm{D}_{\alpha}\right]_{k'k'} + \sum_{k'=1}^{K}\frac{\alpha_{ik'}^{2}}{\left[
\bm{D}_{\alpha}\right]_{k'k'}}\right]
\end{align*}
where $\left[\bm{D}_{\alpha}\right]_{k'k'}$ denotes the $k'$-th diagonal element of
$\bm{D}_{\alpha}$. Considering the case of estimating the $k$-th diagonal element of
$\bm{D}_{\alpha}$ we first separate out the relevant terms
\begin{align*}
-\frac{1}{2}\sum_{i=1}^{M}
\left[\log \prod\mbox{diag}(\bm{D}_{\alpha}) + \sum_{k'=1}^{K}\frac{\alpha_{ik'}^{2}}{\left[
\bm{D}_{\alpha}\right]_{k'k'}}\right] &= \\
-\frac{1}{2}\sum_{i=1}^{M}
\bigg[\log \left[\bm{D}_{\alpha}\right]_{kk} + \frac{\alpha_{ik}^{2}}{\left[\bm{D}_{\alpha}
\right]_{kk}} & + \sum_{k' \ne k}\left[
\log\left[\bm{D}_{\alpha}\right]_{k'k'} + \frac{\alpha_{ik'}^{2}}{\left[\bm{D}_{\alpha}
\right]_{k'k'}}\right] \bigg]
\end{align*}
Then we take the partial derivative with respect to $[\bm{D}_{\alpha}]_{kk}$
\begin{align*}
\frac{ \partial \displaystyle -\frac{1}{2}\sum_{i=1}^{M}
\left[\log \left[\bm{D}_{\alpha}\right]_{kk} + \frac{\alpha_{ik}^{2}}{\left[\bm{D}_{\alpha}
\right]_{kk}}\right]}{\partial [\bm{D}_{\alpha}]_{kk}} &= 
-\frac{1}{2}\sum_{i=1}^{M}
\left[ \frac{1}{[\bm{D}_{\alpha}]_{kk}} - \frac{\alpha_{ik}^{2}}{[\bm{D}_{\alpha}]_{kk}^{2}}
\right]
\end{align*}
Equating to zero and solving for $[\bm{D}_{\alpha}]_{kk}$ gives
\begin{align}
\nonumber -\frac{1}{2}\sum_{i=1}^{M}\left[\frac{1}{[\bm{D}_{\alpha}]_{kk}}\right] &= 
-\frac{1}{2}\sum_{i=1}^{M}\left[\ \frac{\alpha_{ik}^{2}}{[\bm{D}_{\alpha}]_{kk}^{2}}
\right] \\
\nonumber M [\bm{D}_{\alpha}]_{kk} &= \sum_{i=1}^{M} \alpha_{ik}^{2} \\
\widehat{[\bm{D}_{\alpha}]_{kk}} &= \frac{1}{M} \sum_{i=1}^{M} \alpha_{ik}^{2}
\label{eqn:gaussian mlfpca dalpha mle}
\end{align}

\subsection{MLE of \texorpdfstring{$\bm{D}_{\beta_{i}}$}{D_{\beta_{i}}}}
%\subsection{MLE of $D_{\beta_{i}}$}

Following exactly the same procedure to dervive the maximum likelihood estimator of
$\bm{D}_{\beta_{i}}$ gives
\begin{align}
\nonumber -\frac{1}{2}\sum_{j=1}^{n_i}\left[\frac{1}{[\bm{D}_{\beta_i}]_{ll}}\right] &= 
-\frac{1}{2}\sum_{j=1}^{n_i}\left[\ \frac{\beta{ijl}^{2}}{[\bm{D}_{\beta_i}]_{ll}^{2}}
\right] \\
\nonumber n_i [\bm{D}_{\beta_i}]_{ll} &= \sum_{j=1}^{n_i} \beta_{ijl}^{2} \\
\widehat{[\bm{D}_{\beta_i}]_{ll}} &= \frac{1}{n_i} \sum_{j=1}^{n_i} \beta_{ijl}^{2}
\label{eqn:gaussian mlfpca dbeta mle}
\end{align}

\subsection{MLE of $\sigma_{i}^{2}$}

To derive the maximum likelihood estimator of $\sigma_{i}^{2}$ first make the substitution
$\bm{\epsilon}_{i} = \bm{y}_{i} - \bm{B}_{i}\bm{\theta}_{\mu} - \bm{B}_{i}\bm{\Theta}_{\alpha}
\bm{\alpha}_{i} - \widetilde{\bm{B}_{i}}\widetilde{\bm{\Theta}_{\beta_i}}\bm{\beta}_{i}$ for
clarity. Then the relevant terms of (\ref{eqn:multi-level gaussian likelihood}) are $-\frac{1}{2}
\left[N_{i}\log\sigma_{i}^{2} + \sigma_{i}^{-2}\bm{\epsilon}_{i}^{T}\bm{\epsilon}_{i}\right]$.
Taking the partial derivative with respect to $\sigma_{i}^{2}$ gives
\begin{align*}
\frac{ \partial \displaystyle -\frac{1}{2}
\left[N_{i}\log\sigma_{i}^{2} + \sigma_{i}^{-2}\bm{\epsilon}_{i}^{T}\bm{\epsilon}_{i}\right]}
{\partial \sigma_{i}^{2}} &= -\frac{1}{2}
\left[\frac{N_{i}}{\sigma_{i}^{2}} - \frac{1}{\sigma_{i}^{4}}\bm{\epsilon}_{i}^{T}\bm{\epsilon}_{i}
\right]
\end{align*}
Equating to zero and solving for $\sigma_{i}^{2}$ gives
\begin{align}
\widehat{\sigma_{i}^{2}} & = \frac{1}{N_i} \bm{\epsilon}_{i}^{T}\bm{\epsilon}_{i}
\label{eqn:gaussian mlfpca sigma mle}
\end{align}

\subsection{Conditional Expectations}

\begin{table}
\centering
\begin{tabular}{|l|l|}
\hline
Conditional Expectation & Parameters required for \\
\hline
$E[\bm{\alpha}_{i}|\bm{y}_{i}]$ & $\bm{\theta}_{\mu}, \sigma_{i}^{2}$ \\
\hline
$E[\bm{\beta}_{i}|\bm{y}_{i}]$ & $\bm{\theta}_{\mu}, \sigma_{i}^{2}$ \\
\hline
$E[\bm{\alpha}_{i}\bm{\alpha}_{i}^{T}|\bm{y}_{i}]$ &
  $\bm{\Theta}_{\alpha}, \bm{D}_{\alpha}$ \\
\hline
$E[\bm{\beta}_{i}\bm{\beta}_{i}^{T}|\bm{y}_{i}]$ &
  $\bm{\Theta}_{\beta_{i}}, \bm{D}_{\beta_i}$ \\
\hline
$E[\bm{\alpha}_{i}\bm{\beta}_{i}^{T}|\bm{y}_{i}]$ &
  $\bm{\Theta}_{\alpha}, \bm{\Theta}_{\beta_{i}}$ \\
\hline
$E[\bm{\epsilon}_{i}^{T}\bm{\epsilon}_{i}|\bm{y}_{i}]$ &
  $\sigma_{i}^{2}$ \\
\hline
\end{tabular}
\caption{The required conditional expectations for the EM algorithm for the Gaussian multi-level
reduced-rank fPCA model}
\label{table:conditional expectations gaussian mlfpca}
\end{table}

\begin{table}
\centering
\begin{tabular}{|l|l|}
\hline
Conditional Expectation & Parameters required for \\
\hline
$E[\bm{\alpha}_{i}|\bm{y}_{i}]$ & $\bm{\theta}_{\mu}, \sigma_{i}^{2}, \xi_{\alpha_k},
\sigma_{\alpha_k}^{2}, \lambda_{\alpha_k}, \nu_{\alpha_k}$ \\
\hline
$E[\bm{\beta}_{i}|\bm{y}_{i}]$ & $\bm{\theta}_{\mu}, \sigma_{i}^{2}$ \\
\hline
$E[\bm{\alpha}_{i}\bm{\alpha}_{i}^{T}|\bm{y}_{i}]$ &
  $\bm{\Theta}_{\alpha}$ \\
\hline
$E[\bm{\beta}_{i}\bm{\beta}_{i}^{T}|\bm{y}_{i}]$ &
  $\bm{\Theta}_{\beta_{i}}, \bm{D}_{\beta_i}$ \\
\hline
$E[\bm{\alpha}_{i}\bm{\beta}_{i}^{T}|\bm{y}_{i}]$ &
  $\bm{\Theta}_{\alpha}, \bm{\Theta}_{\beta_{i}}$ \\
\hline
$E[\bm{\epsilon}_{i}^{T}\bm{\epsilon}_{i}|\bm{y}_{i}]$ &
  $\sigma_{i}^{2}$ \\
\hline
\end{tabular}
\caption{The required conditional expectations for the EM algorithm for the skew-$t$-normal
multi-level reduced-rank FPCA model}
\label{table:conditional expectations gaussian mlfpca}
\end{table}

Supplementary Table \ref{table:conditional expectations gaussian mlfpca} summarises the requried conditional
expectations for the EM algorithm, based on the sufficient statistics of the maximum likelihood
estimators derived above. In order to derive these conditional expectations we first write
\begin{multline*}
\left[\begin{array}{c}\bm{\alpha}_{i} \\ \bm{\beta}_{i} \\ \bm{\epsilon}_{i} \\ 
\bm{y}_{i} \end{array}\right] \sim
MVN\left( \left[\begin{array}{c}\bm{0} \\ \bm{0} \\ \bm{0} \\ \bm{B}_{i}\bm{\theta}_{\mu}
\end{array}\right]\right. , \\ \left.\left[\begin{array}{cccc}\bm{D}_{\alpha} & \bm{0} & \bm{0} &
\bm{D}_{\alpha}\bm{\Theta}_{\alpha}^{T}\bm{B}_{i}^{T} \\
\bm{0} & \widetilde{\bm{D}_{\beta_i}} & \bm{0} & \widetilde{\bm{D}_{\beta_i}}
\widetilde{\bm{\Theta}_{\beta_i}^{T}}\widetilde{\bm{B}_{i}^{T}} \\
\bm{0} & \bm{0} & \sigma_{i}^{2}\bm{I}_{N_i \times N_i} & \sigma_{i}^{2}\bm{I}_{N_i \times N_i}  \\
\bm{B}_{i}\bm{\Theta}_{\alpha}\bm{D}_{\alpha} & \widetilde{\bm{B}_{i}}
\widetilde{\bm{\Theta}_{\beta_i}}\widetilde{\bm{D}_{\beta_i}} & 
\sigma_{i}^{2}\bm{I}_{N_i \times N_i} & \bm{V}_{i}\end{array}\right] \right)
\end{multline*}
where recall that $\bm{V}_{i}$ is given in (\ref{eqn:gaussian mlfpca Vi}). Using the standard result
given in \cite{Anderson1958} we can then write
\begin{align*}
%\bm{\alpha}_{i} | \bm{y}_{i} \sim MVN\bigg(\bm{D}_{\alpha}\bm{\Theta}_{\alpha}^{T}\bm{B}_{i}^{T}
%\bm{V}_{i}^{-1}(\bm{y}_{i} - &\bm{B}_{i}\bm{\theta}_{\mu}), \\
%\bm{D}_{\alpha} - 
%\bm{D}_{\alpha}\bm{\Theta}_{\alpha}^{T}\bm{B}_{i}^{T}&\bm{V}_{i}^{-1}
%\bm{B}_{i}\bm{\Theta}_{\alpha}\bm{D}_{\alpha}\bigg) \\
%\bm{\beta}_{i} | \bm{y}_{i} \sim MVN\bigg(\widetilde{\bm{D}_{\beta_i}}
%\widetilde{\bm{\Theta}_{\beta_i}^{T}}\widetilde{\bm{B}_{i}^{T}}
%\bm{V}_{i}^{-1}(\bm{y}_{i} - &\bm{B}_{i}\bm{\theta}_{\mu}), \\
%\widetilde{\bm{D}_{\beta_i}} - 
%\widetilde{\bm{D}_{\beta_i}}
%\widetilde{\bm{\Theta}_{\beta_i}^{T}}\widetilde{\bm{B}_{i}^{T}}&\bm{V}_{i}^{-1}
%\widetilde{\bm{B}_{i}}
%\widetilde{\bm{\Theta}_{\beta_i}}\widetilde{\bm{D}_{\beta_i}}\bigg) \\
\left[\left.\begin{array}{c}\bm{\alpha}_{i} \\ \bm{\beta}_{i}\end{array}\right]\right|\bm{y}_{i}
\sim MVN\bigg(\left[\begin{array}{c}\bm{D}_{\alpha}\bm{\Theta}_{\alpha}^{T}\bm{B}_{i}^{T}
\\ \widetilde{\bm{D}_{\beta_i}}
\widetilde{\bm{\Theta}_{\beta_i}^{T}}\widetilde{\bm{B}_{i}^{T}}
\end{array}\right] \bm{V}_{i}^{-1}(\bm{y}_{i} - &\bm{B}_{i}\bm{\theta}_{\mu}) ,
\\
\left[\begin{array}{cc}\bm{D}_{\alpha_i} & \bm{0} \\ \bm{0} & \bm{D}_{\beta_{i}}\end{array}\right]
- \left[\begin{array}{c}\bm{D}_{\alpha}\bm{\Theta}_{\alpha}^{T}\bm{B}_{i}^{T}
\\ \widetilde{\bm{D}_{\beta_i}}
\widetilde{\bm{\Theta}_{\beta_i}^{T}}\widetilde{\bm{B}_{i}^{T}}
\end{array}\right] \bm{V}_{i}^{-1}  & \left[\begin{array}{cc}
\bm{B}_{i}\bm{\Theta}_{\alpha}\bm{D}_{\alpha} & \widetilde{\bm{B}_{i}}
\widetilde{\bm{\Theta}_{\beta_i}}\widetilde{\bm{D}_{\beta_i}}\end{array}\right]\bigg) \\
\bm{\epsilon}_{i} | \bm{y}_{i} \sim MVN\bigg(\sigma^{2}\bm{V}_{i}^{-1}
(\bm{y}_{i} - \bm{B}_{i}\bm{\theta}_{\mu})&, \sigma^{2}\bm{I}_{N_{i} \times N_i} -
\sigma^{4}\bm{V}_{i}^{-1}\bigg)
\end{align*}
Hence we immediately have
\begin{align}
\label{eqn:gaussian mlfpca alpha ce}
E[\bm{\alpha}_{i}|\bm{y}_{i}] &= \widehat{\bm{\alpha}_{i}} =
\bm{D}_{\alpha}\bm{\Theta}_{\alpha}^{T}\bm{B}_{i}^{T}
\bm{V}_{i}^{-1}(\bm{y}_{i} - \bm{B}_{i}\bm{\theta}_{\mu}) \\
E[\bm{\beta}_{i}|\bm{y}_{i}] &= \widehat{\bm{\beta}_{i}} = \widetilde{\bm{D}_{\beta_i}}
\widetilde{\bm{\Theta}_{\beta_i}^{T}}\widetilde{\bm{B}_{i}^{T}}
\bm{V}_{i}^{-1}(\bm{y}_{i} - \bm{B}_{i}\bm{\theta}_{\mu})
\label{eqn:gaussian mlfpca beta ce}
\end{align}
and the remaining results follow from the definition of covariance
\begin{align}
\label{eqn:gaussian mlfpca alphaalpha ce}
E[\bm{\alpha}_{i}\bm{\alpha}_{i}^{T}|\bm{y}_{i}] &= \widehat{\bm{\alpha}_{i}\bm{\alpha}_{i}^{T}} =
\widehat{\bm{\alpha}_{i}}
\widehat{\bm{\alpha}_{i}}^{T} + \bm{D}_{\alpha} - 
\bm{D}_{\alpha}\bm{\Theta}_{\alpha}^{T}\bm{B}_{i}^{T}\bm{V}_{i}^{-1}
\bm{B}_{i}\bm{\Theta}_{\alpha}\bm{D}_{\alpha} \\
\label{eqn:gaussian mlfpca betabeta ce} E[\bm{\beta}_{i}\bm{\beta}_{i}^{T}|\bm{y}_{i}] &=
\widehat{\bm{\beta}_{i}\bm{\beta}_{i}^{T}} =
\widehat{\bm{\beta}_{i}}\widehat{\bm{\beta}_{i}}^{T} + \widetilde{\bm{D}_{\beta_i}} - 
\widetilde{\bm{D}_{\beta_i}}
\widetilde{\bm{\Theta}_{\beta_i}^{T}}\widetilde{\bm{B}_{i}^{T}}\bm{V}_{i}^{-1}
\widetilde{\bm{B}_{i}}
\widetilde{\bm{\Theta}_{\beta_i}}\widetilde{\bm{D}_{\beta_i}} \\
%\nonumber
E[\bm{\alpha}_{i}\bm{\beta}_{i}^{T}|\bm{y}_{i}] &= \widehat{\bm{\alpha}_{i}
\bm{\beta}_{i}^{T}} =
\widehat{\bm{\alpha}_{i}}\widehat{\bm{\beta}_{i}}^{T} %+
%\left[\begin{array}{cc}\bm{D}_{\alpha_i} & \bm{0} \\ \bm{0} & \bm{D}_{\beta_{i}}\end{array}\right]
- %\\
%& \left[\begin{array}{c}\bm{D}_{\alpha}\bm{\Theta}_{\alpha}^{T}\bm{B}_{i}^{T}
%\\ \widetilde{\bm{D}_{\beta_i}}
%\widetilde{\bm{\Theta}_{\beta_i}^{T}}\widetilde{\bm{B}_{i}^{T}}
%\end{array}\right]
\bm{D}_{\alpha}\bm{\Theta}_{\alpha}^{T}\bm{B}_{i}^{T}
\bm{V}_{i}^{-1}\widetilde{\bm{B}_{i}}\widetilde{\bm{\Theta}_{\beta_i}}\widetilde{\bm{D}_{\beta_i}}
%\left[\begin{array}{cc}
%\bm{B}_{i}\bm{\Theta}_{\alpha}\bm{D}_{\alpha} & \widetilde{\bm{B}_{i}}
%\widetilde{\bm{\Theta}_{\beta_i}}\widetilde{\bm{D}_{\beta_i}}\end{array}\right]
\label{eqn:gaussian mlfpca alphabeta ce}
\end{align}
For the case of $E[\bm{\epsilon}_{i}^{T}\bm{\epsilon}_{i}|\bm{y}_{i}]$ first note that
$E[\bm{\epsilon}_{i}^{T}\bm{\epsilon}_{i}|\bm{y}_{i}] = tr(
E[\bm{\epsilon}_{i}\bm{\epsilon}_{i}^{T}|\bm{y}_{i}])$. Then, letting $\widehat{\bm{\epsilon}_{i}}
= \bm{y}_{i} - \bm{B}_{i}\widehat{\bm{\theta}_{\mu}} - \bm{B}_{i}\widehat{\bm{\Theta}_{\alpha}}
\widehat{\bm{\alpha}_{i}} - \widetilde{\bm{B}_{i}}\widetilde{\widehat{\bm{\Theta}_{\beta_i}}}
\widehat{\bm{\beta}_{i}}$ and from the definition of covariance we have
\begin{align*}
E[\bm{\epsilon}_{i}\bm{\epsilon}_{i}^{T}|\bm{y}_{i}] &= \widehat{\bm{\epsilon}_{i}
\bm{\epsilon}_{i}^{T}} = \widehat{\bm{\epsilon}_{i}}\widehat{\bm{\epsilon}_{i}}^{T} + \sigma_i^{2}
\bm{I}_{N_i \times N_i} - \sigma_i^{4}\bm{V}_{i}^{-1}
\end{align*}
so that
\begin{align}
\nonumber E[\bm{\epsilon}_{i}^{T}\bm{\epsilon}_{i}|\bm{y}_{i}] = \widehat{\bm{\epsilon}_{i}^{T}
\bm{\epsilon}_{i}} &= \mbox{trace}(\widehat{\bm{\epsilon}_{i}}\widehat{\bm{\epsilon}_{i}}^{T} +
\sigma_i^{2}\bm{I}_{N_i \times N_i} - \sigma_i^{4}\bm{V}_{i}^{-1}) \\
&= \widehat{\bm{\epsilon}_{i}}^{T}\widehat{\bm{\epsilon}_{i}} + N_i\sigma_i^2 - \sigma_i^4
tr(\bm{V}_{i}^{-1})
\label{eqn:gaussian mlfpca epsilonepsilon ce}
\end{align}

\section{Deriving the posterior distributions of $\gamma_{ik}$ and $\tau_{ik}$}

Recall that the skew-$t$-normal density is given by
\begin{align}\label{eqn:skew-t-normal density}
f(\alpha_{ik}|\xi_{\alpha_k},\sigma^2_{\alpha_k},\lambda_{\alpha_k},\nu_{\alpha_k}) =
  2t_{\nu_{\alpha_k}}(\alpha_k;\xi_{\alpha_k},\sigma^2_{\alpha_k})\Phi\left(\frac{\alpha_{ik} -
    \xi_{\alpha_k}}{\sigma_{\alpha_k}}\lambda_{\alpha_k}
\right)
\end{align}
The joint density $f(\alpha_{ik},\gamma_{ik},\tau_{ik})$ is given by
\begin{multline}\label{eqn:joint density skew-t-normal}
f(\alpha_{ik},\gamma_{ik},\tau_{ik}) =
f(\alpha_{ik} | \gamma_{ik},\tau_{ik})f(\gamma_{ik}| \tau_{ik}) f(\tau_{ik}) \\
= \frac{\sqrt{\tau_{ik} + \lambda_{\alpha_{k}}^{2}}}{\sqrt{2\pi}
\sigma_{\alpha_{k}}}
 \exp \left\{ -\frac{\tau_{ik} + \lambda_{\alpha_k}^{2}}{2\sigma_{\alpha_k}^{2}}\left(\alpha_{ik} -
\xi_{\alpha_k} - \frac{\sigma_{\alpha_{k}}\lambda_{\alpha_k}}{\tau_{ik} + 
\lambda_{\alpha_k}^{2}}\gamma_{\alpha_k}\right)^{2}\right\} \times \\
I(\gamma_{ik} > 0) \frac{2\sqrt{\tau_{ik}}}{\sqrt{2\pi}\sqrt{\tau_{ik} +
\lambda_{\alpha_k}^{2}}} \exp\left\{-\frac{\tau_{ik}}{2(\tau_{ik} + \lambda_{\alpha_k}^{2})}
\gamma_{ik}^{2}\right\} \times \\
\left[\frac{\nu_{\alpha_k}}{2}\right]^{\nu_{\alpha_k}/2}\tau_{ik}^{\nu_{\alpha_k}/2 - 1}
\frac{\exp\left\{- \displaystyle \frac{\nu_{\alpha_k}}{2}\tau_{ik}\right\}}
{\Gamma\left(\displaystyle \frac{\nu_{\alpha_k}}{2}\right)}
\end{multline}
We first integrate
$\gamma_{ik}$ out of the joint density $f(\alpha_{ik},\tau_{ik},\gamma_{ik})$ to obtain
\begin{multline*}
\int_{-\infty}^{\infty} f(\alpha_{ik},\gamma_{ik},\tau_{ik})d\gamma_{ik} =
\frac{1}{\pi\sigma_{\alpha_k}}\frac{(\nu_{\alpha_k}/2)^
{\nu_{\alpha_k}/2}}{\Gamma(\nu_{\alpha_k} / 2)}\tau_{ik}^{(\nu_{\alpha_k}+1)/2 - 1}
\exp\left\{- \displaystyle \frac{\nu_{\alpha_k}}{2}\tau_{ik}\right\} \times \\
\exp\left\{-\frac{1}{2}
\frac{\tau_{ik}}{\sigma_{\alpha_k}^{2}}(\alpha_{ik} - \xi_{\alpha_k})^{2} \right\} \times  \\
\int_{-\infty}^{\infty}
I(\gamma_{ik} > 0)
\exp\left\{- \frac{1}{2} \left(\gamma_{ik} - (\alpha_{ik} - \xi_{\alpha_k})\frac{\lambda_{\alpha_k}}
{\sigma_{\alpha_k}}\right)^{2}\right\}d\gamma_{ik} \\
= \frac{1}{\pi\sigma_{\alpha_k}}\frac{(\nu_{\alpha_k}/2)^
{\nu_{\alpha_k}/2}}{\Gamma(\nu_{\alpha_k} / 2)}\tau_{ik}^{(\nu_{\alpha_k}+1)/2 - 1}
\exp\left\{- \displaystyle \frac{\nu_{\alpha_k}}{2}\tau_{ik}\right\} \exp\left\{-\frac{1}{2}
\frac{\tau_{ik}}{\sigma_{\alpha_k}^{2}}(\alpha_{ik} - \xi_{\alpha_k})^{2} \right\} \times  \\
\int_{-\infty}^{0}
\exp\left\{- \frac{1}{2} \left(\gamma_{ik} - (\alpha_{ik} - \xi_{\alpha_k})
\frac{\lambda_{\alpha_k}}{\sigma_{\alpha_k}}\right)^{2}\right\}d\gamma_{ik}
\end{multline*}
Making the substitution $x = \gamma_{ik} - (\alpha_{ik} - \xi_{\alpha_k})
\frac{\lambda_{\alpha_k}}{\sigma_{\alpha_k}}$, the integration becomes
\begin{multline*}
\int_{-\infty}^{0}
\exp\left\{- \frac{1}{2} \left(\gamma_{ik} - (\alpha_{ik} - \xi_{\alpha_k})
\frac{\lambda_{\alpha_k}}{\sigma_{\alpha_k}}\right)^{2}\right\}d\gamma_{ik} = \\
\int_{-\infty}^{(\alpha_{ik} - \xi_{\alpha_k})
\frac{\lambda_{\alpha_k}}{\sigma_{\alpha_k}}}
\exp\left\{- \frac{1}{2}x^{2}\right\}dx
 = \sqrt{2\pi}\Phi\left((\alpha_{ik} - \xi_{\alpha_k})
\frac{\lambda_{\alpha_k}}{\sigma_{\alpha_k}}\right)
\end{multline*}
where $\Phi(\cdot)$ denotes the cdf of the standard normal distribution. Hence we have
\begin{multline}
\label{eqn:joint density2}
f(\alpha_{ik},\tau_{ik}) = \left(\frac{2}{\pi\sigma_{\alpha_k}^{2}}\right)^{1/2}
\frac{(\nu_{\alpha_k}/2)^
{\nu_{\alpha_k}/2}}{\Gamma(\nu_{\alpha_k} / 2)}\tau_{ik}^{(\nu_{\alpha_k}+1)/2 - 1}
\exp\left\{- \displaystyle \frac{\nu_{\alpha_k}}{2}\tau_{ik}\right\} \times
\\ \exp\left\{-\frac{1}{2}
\frac{\tau_{ik}}{\sigma_{\alpha_k}^{2}}(\alpha_{ik} - \xi_{\alpha_k})^{2} \right\} \times
\Phi\left((\alpha_{ik} - \xi_{\alpha_k})
\frac{\lambda_{\alpha_k}}{\sigma_{\alpha_k}}\right)
\end{multline}
Dividing (\ref{eqn:joint density skew-t-normal}) by (\ref{eqn:joint density2}) gives
$f(\gamma_{ik}|\alpha_{ik},\tau_{ik})$ as
\begin{equation*}
f(\gamma_{ik}|\alpha_{ik},\tau_{ik}) = \frac{I(\gamma_{ik} > 0)}
{\Phi\left((\alpha_{ik} - \xi_{\alpha_k})\frac{\lambda_{\alpha_k}}{\sigma_{\alpha_k}}\right)
\sqrt{2\pi}}\exp\left\{ - \frac{1}{2} \left(\gamma_{ik} - (\alpha_{ik} - \xi_{\alpha_k})
\frac{\lambda_{\alpha_k}}{\sigma_{\alpha_k}}\right)^{2} \right\}
\end{equation*}
As $\tau_{ik}$ does not appear anywhere on the right hand side, it implies that, conditional on
$\alpha_{ik}$, $\gamma_{ik}$ and $\tau_{ik}$ are independent. Hence
\begin{equation}
\label{eqn:gamma conditional}
f(\gamma_{ik}|\alpha_{ik}) = \frac{I(\gamma_{ik} > 0)}
{\Phi\left((\alpha_{ik} - \xi_{\alpha_k})\frac{\lambda_{\alpha_k}}{\sigma_{\alpha_k}}\right)
\sqrt{2\pi}}\exp\left\{ - \frac{1}{2} \left(\gamma_{ik} - (\alpha_{ik} - \xi_{\alpha_k})
\frac{\lambda_{\alpha_k}}{\sigma_{\alpha_k}}\right)^{2} \right\}
\end{equation}
From (\ref{eqn:gamma conditional}) it follows that $\gamma_{ik}|\alpha_{ik} \sim TN(
(\alpha_{ik} - \xi_{\alpha_k})\frac{\lambda_{\alpha_k}}{\sigma_{\alpha_k}},1;(0,\infty))$ and so
\begin{equation}\label{eqn:gamma conditional expectation}
E[\gamma_{ik}|\alpha_{ik}] = (\alpha_{ik} - \xi_{\alpha_k})
\frac{\lambda_{\alpha_k}}{\sigma_{\alpha_k}} + \frac{\phi\left((\alpha_{ik} - \xi_{\alpha_k})
\frac{\lambda_{\alpha_k}}{\sigma_{\alpha_k}} \right)}{\Phi\left((\alpha_{ik} - \xi_{\alpha_k})
\frac{\lambda_{\alpha_k}}{\sigma_{\alpha_k}} \right)}
\end{equation}
For $f(\tau_{ik}|\alpha_{ik})$, we divide (\ref{eqn:joint density2}) by
(\ref{eqn:skew-t-normal density}) yielding
\begin{multline}\label{eqn:tau conditional}
f(\tau_{ik}|\alpha_{ik}) = \frac{1}{\Gamma((\nu_{\alpha_k}+1)/2)}\left(\frac{\nu_{\alpha_k}
+ (\alpha_{ik} - \xi_{\alpha_k})^{2}}{2\sigma_{\alpha_k}^2}\right)^{(\nu_{\alpha_k}+1)/2}
\tau_{ik}^{(\nu_{\alpha_k}+1)/2-1} \times \\
\exp\left\{-\frac{\tau_{ik}\left[\nu_{\alpha_k} + (\alpha_{ik} -
\xi_{\alpha_k})^2\right]}{2\sigma^2_{\alpha_k}}\right\}
\end{multline}
and so
\begin{equation*}
\tau_{ik}|\alpha_{ik} \sim \Gamma \left(\frac{\nu_{\alpha_k}+1}{2},
\frac{\nu_{\alpha_k} + (\alpha_{ik} - \xi_{\alpha_k})^{2}/\sigma_{\alpha_k}^{2}}{2}\right)
\end{equation*}

\section{Distribution of \texorpdfstring{$f(\bm{\alpha}_{i},\bm{\beta}_{i}|\bm{y}_{i},\bm{\tau}_{i},\bm{\gamma}_{i})$}{$f(\alpha_{i},\beta_{i}|y_{i},\tau_{i},\gamma_{i})$}}
\begin{multline*}
\left.\left[\begin{array}{c}\bm{\alpha}_{i} \\ \bm{\beta}_{i} \end{array}\right]
\right|\bm{y}_{i},\bm{\tau}_{i},\bm{\gamma}_{i} \sim MVN\Bigg(
\\\left[\begin{array}{c}
\bm{\mu}_{\alpha} \\ \bm{0}\end{array}\right] + \left[\begin{array}{c}
\mbox{diag}(\bm{v}_{\alpha})\bm{\Theta}_{\alpha}^{T}\bm{B}_{i}^{T} \\
\widetilde{\bm{D}_{\beta_i}}\widetilde{\bm{\Theta}_{\beta_i}^{T}}\widetilde{\bm{B}_{i}^{T}}
\end{array}\right]\bm{V}_{y_i|\tau_i,\gamma_i}^{-1}(\bm{y}_{i} - \bm{B}_{i}\bm{\theta}_{\mu} -
\bm{B}_{i}\bm{\Theta}_{\alpha}\bm{\alpha}_{i}),
\\
\left[\begin{array}{cc}\mbox{diag}(\bm{v}_{\alpha}) & \bm{0} \\ \bm{0} &
\widetilde{\bm{D}_{\beta_i}}\end{array}\right] - \\
\left[\begin{array}{c}\mbox{diag}(\bm{v}_{\alpha})\bm{\Theta}_{\alpha}^{T}\bm{B}_{i}^{T} \\
\widetilde{\bm{D}_{\beta_i}}\widetilde{\bm{\Theta}_{\beta_i}^{T}}\widetilde{\bm{B}_{i}^{T}}
\end{array}\right]\bm{V}_{y_i|\tau_i,\gamma_i}^{-1}\left[\begin{array}{cc}
\bm{B}_{i}\bm{\Theta}_{\alpha}\mbox{diag}(\bm{v}_{\alpha}) & 
\widetilde{\bm{B}_{i}}\widetilde{\bm{\Theta}_{\beta_i}}\widetilde{\bm{D}_{\beta_i}}
\end{array}\right]\Bigg)
\end{multline*}

\section{EM algorithm initialisation}
To obtain initial estimates of the parameters, the following procedure is adopted, with similar
approaches used elsewhere in the literature \citep{Peng2009,Zhou2010}.

First a spline is fit to the entire data, ignoring variable and replicate labels, using least
squares in order to obtain initial values for $\widehat{\bm{\theta}_{\mu}}$. Next, the grand mean is
subtracted from all observations and each variable is fit independently with a spline using least
squares. With $M$ variables in the data set and a $p$-dimensional spline basis, this results in an
$M \times p$ matrix of spline basis coefficients. A standard PCA is performed on this matrix
yielding $p$ eigenvectors each of length $p$. The first $K$ eigenvectors are taken to be the
initial $\widehat{\bm{\Theta}_{\alpha}}$ matrix, and skew-$t$-normal distributions are fit to the
loadings in order to initialise $\hat{\xi}_{\alpha_k}$, $\hat{\sigma}_{\alpha_k}^2$,
$\hat{\lambda}_{\alpha_k}$ and $\hat{\nu}_{\alpha_k}$. Next the variable means are subtracted from
the observations and each replicate for each variable is fit independently with a spline using least
squares. Note that it will be necessary to instead perform a ridge regression in those cases where
the replicate has not been observed at all time points as the matrix $\bm{B}_{ij}^{T}\bm{B}_{ij}$
will not be of full rank. In fact, we suggest to perform a ridge regression in all instances, even
when the matrix $\bm{B}_{ij}^{T}\bm{B}_{ij}$ is of full rank, as
this imposes some smoothness and avoids overfitting the data which may lead to unrealistically small
initial values for $\widehat{\sigma_{i}^{2}}$. As a result of this procedure, an $n_{i} \times p$
matrix of spline basis coefficients is obtained for each variable. As before, a standard PCA is
performed on this matrix and the first $L_{i}$ eigenvectors retained as the initial values for
$\widehat{\bm{\Theta}_{\beta_i}}$. The corresponding eigenvalues are used for the initial values
for the diagonal element of $\widehat{\bm{D}_{\beta_i}}$. After subtracting these replicate curves
from the observations we are left with initial estimates $\widehat{\bm{\epsilon}_{i}}$ and the
sample variance of these residuals can be used as initial values for $\widehat{\sigma_{i}^{2}}$ for
each variable.

\begin{figure}[htbp]
\includegraphics[width=\textwidth]{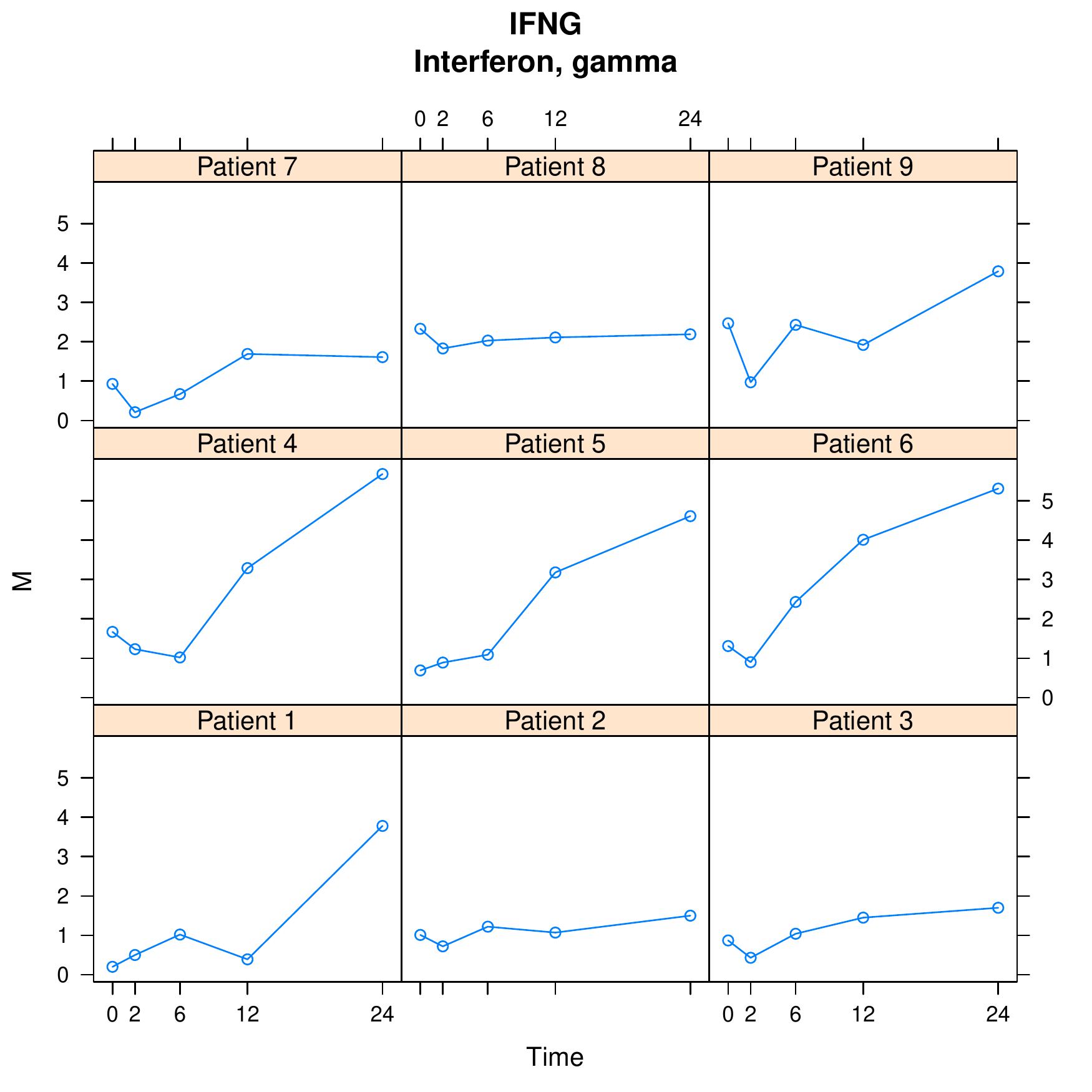}
\caption
{Raw data for expression levels of interferon-gamma measured in blood samples from nine
human patients to which the BCG vaccine for tuberculosis was added. Interferon-gamma is well-known
to play an important role in the immune response to tuberculosis infection. Note the heterogeneous
response with patients two, three, seven and eight exhibiting flat profiles, while the other
patients end the time course with elevated levels of gene expression. Compare this to the much more
homogeneous profiles for tumour necrosis factor-alpha given in Supplementary Figure
\ref{fig:TNF raw data}}
\label{fig:IFNG raw data}
\end{figure}
\begin{figure}[htbp]
\includegraphics[width=\textwidth]{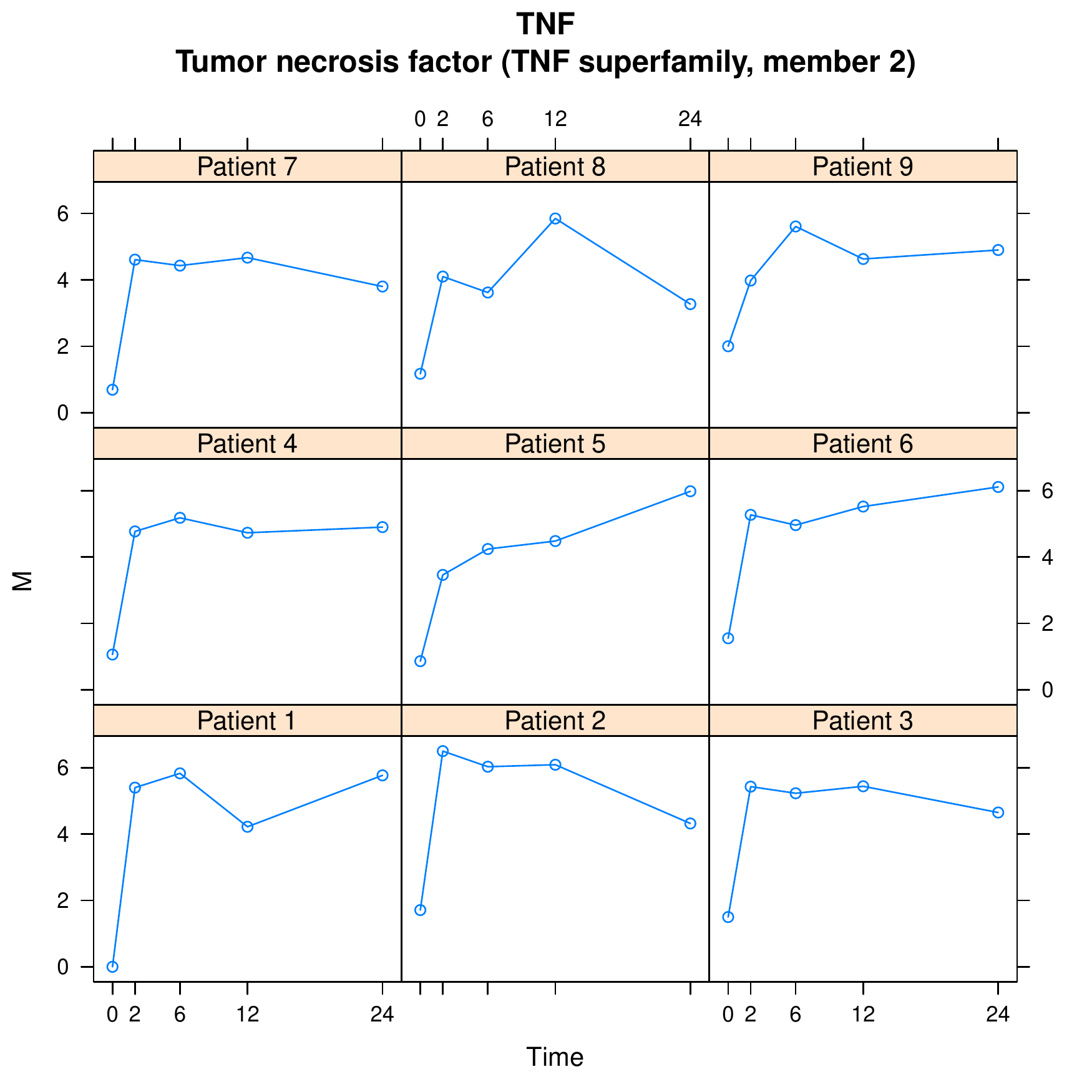}
\caption
{Raw data for expression levels of tumour necrosis factor-alpha measured in blood samples
from nine human patients to which the BCG vaccine for tuberculosis was added. Tumour necrosis
factor-alpha is well-known to be associated with tuberculosis infection. Note the subtle differences
between the profiles such as the expression level at which the time course begins, the level to
which they peak, and whether they plateau, continue to rise or start to fall by the end of
the time course. However, compared with the data given in Supplementary
Figure \ref{fig:IFNG raw data} for
interferon-gamma, these profiles are much more homogeneous across the different patients.}
\label{fig:TNF raw data}
\end{figure}
\begin{figure}[p]
\includegraphics[width=\textwidth]{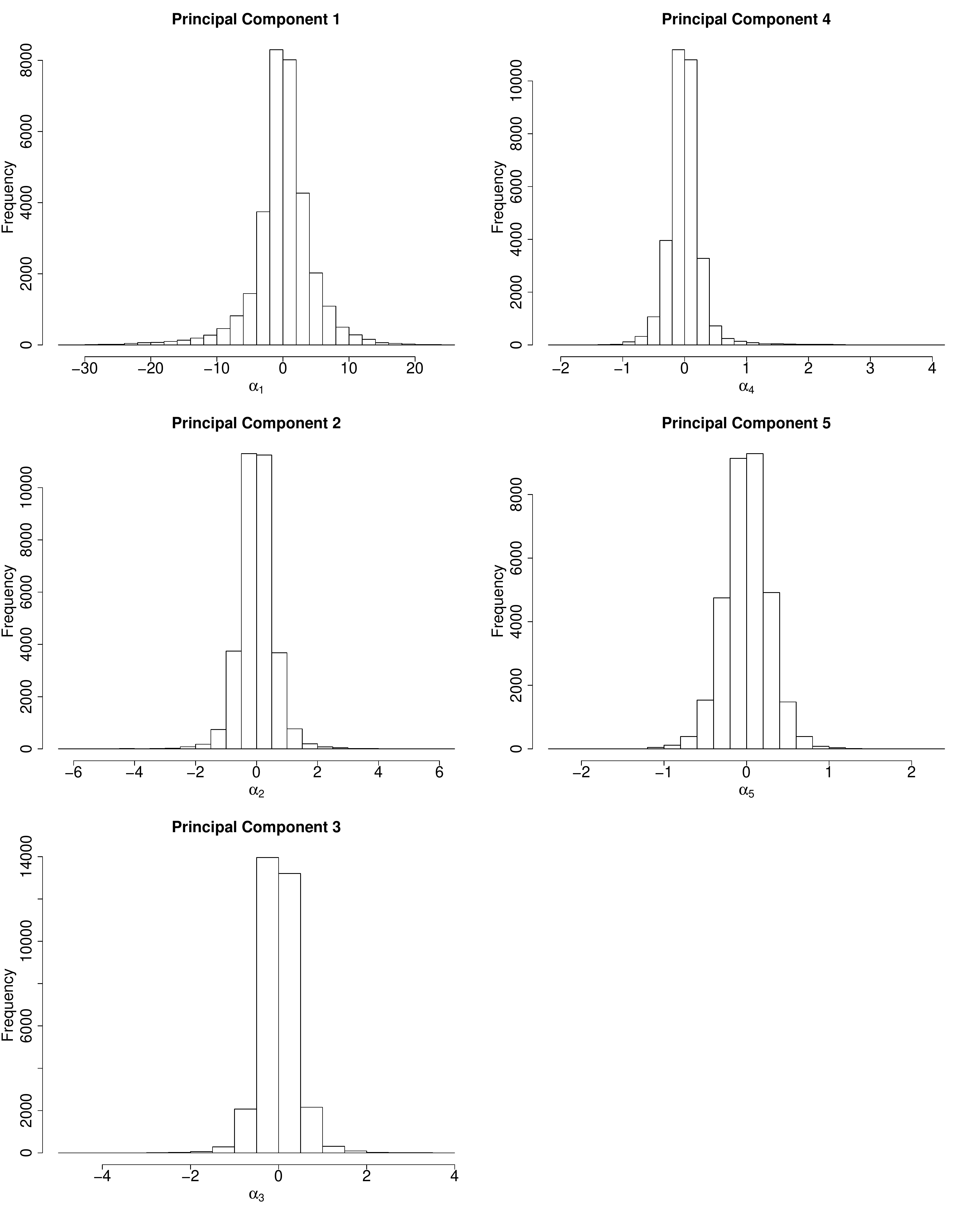}
\caption
{\label{fig:bcg initial loadings}Initialised variable-level loadings for a genomics data set
studying the genetic response to BCG infection. Note the departure from normality in all
instances with many outliers and varying levels of skewness.}
\end{figure}
\begin{figure}[p]
\includegraphics[width=\textwidth]{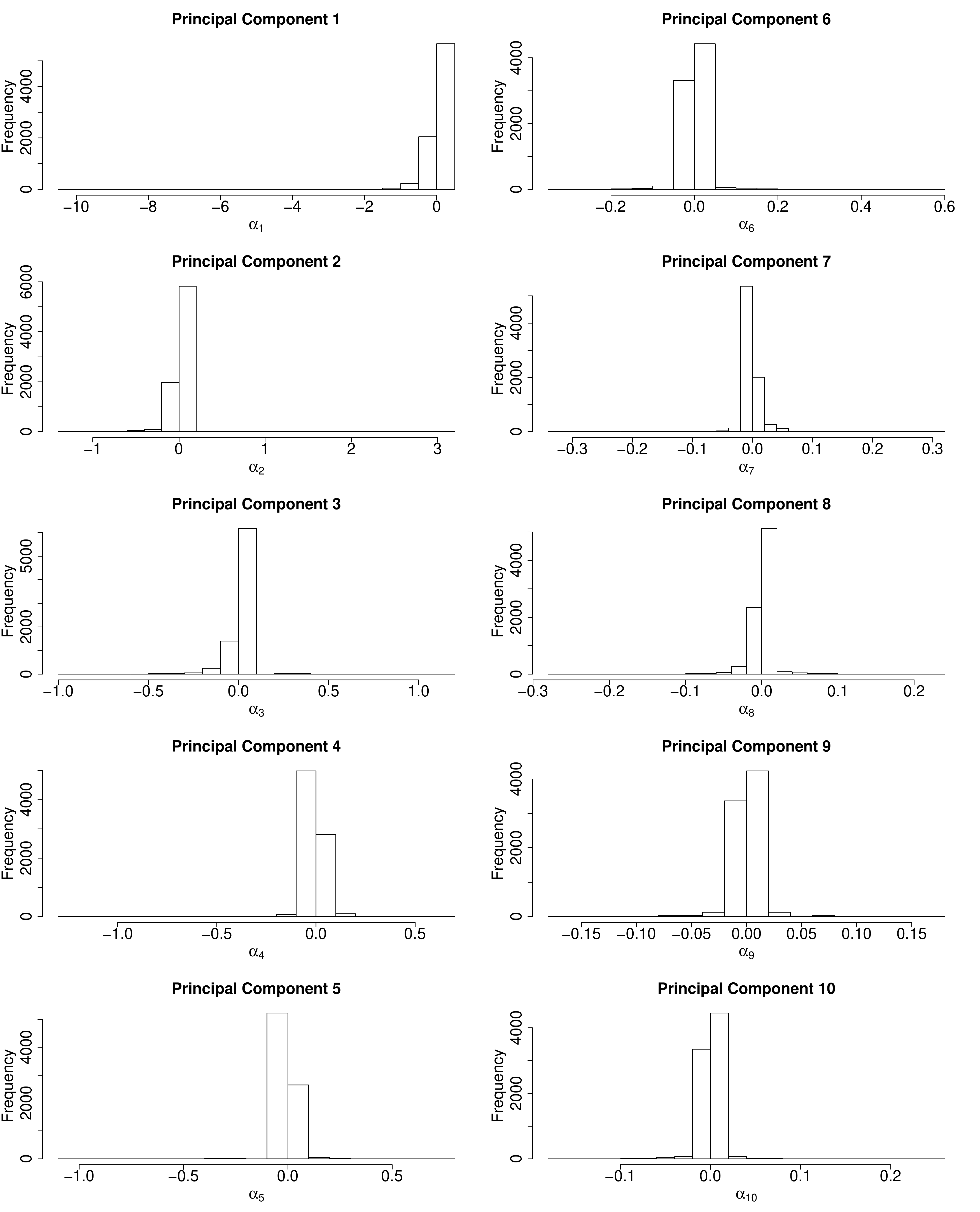}
\caption
{\label{fig:tim initial loadings}Initialised variable-level loadings from a metabolomics
toxicology study. Note the departure from normality in all instances with many outliers and varying
levels of skewness. The extreme skew for the loadings on the first principal component, such that
the distribution is essentially a truncated-\textit{t} distribution, is a result of the fact that
there are no negative observations in this data set.}
\end{figure}
% \begin{figure}[p]
% \includegraphics[width=\textwidth]{mlfpcaSimulationGrandMean.pdf}
% \caption[Simulated grand mean function used in our
% simulation study to compare the Gaussian single- and multi-level reduced-rank fPCA models]
% {\label{fig:mlfpca simulation grand mean}Simulated grand mean function used in our
% simulation study to compare the Gaussian single- and multi-level reduced-rank fPCA models. The shape
% of this curve is motivated by real biological data.}
% \end{figure}
\begin{figure}[tb]
\includegraphics[width=\textwidth]{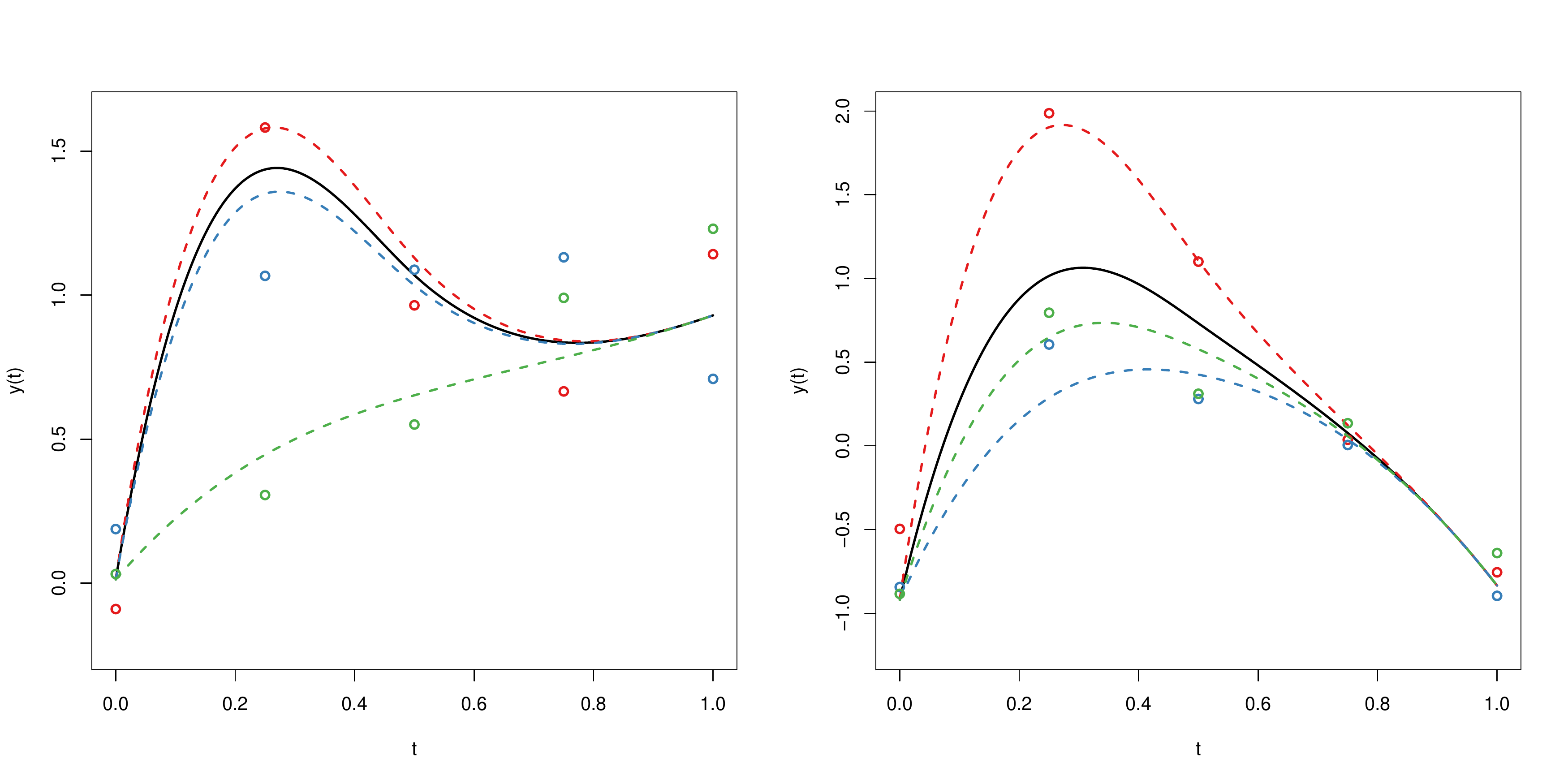}
\caption[Example simulated data for two variables produced
under the simulation setting for the Gaussian multi-level reduced-rank fPCA model]
{\label{fig:mlfpca simulation examples}Example simulated data for two variables produced
under the simulation setting for the Gaussian multi-level reduced-rank fPCA model.
The solid black lines correspond to the variable mean curve. The coloured dashed lines are the
individual replicate curves. For clarity only three replicates are shown. Final observations
including simulated noise are shown as circles, and are also colour-coded for replicate. Note how
the combination of the two variable-level and the single replicate-level principal component
functions described above result in a wide range of replicate curves.}
\end{figure}
\begin{figure}[p]
\includegraphics[width=\textwidth]{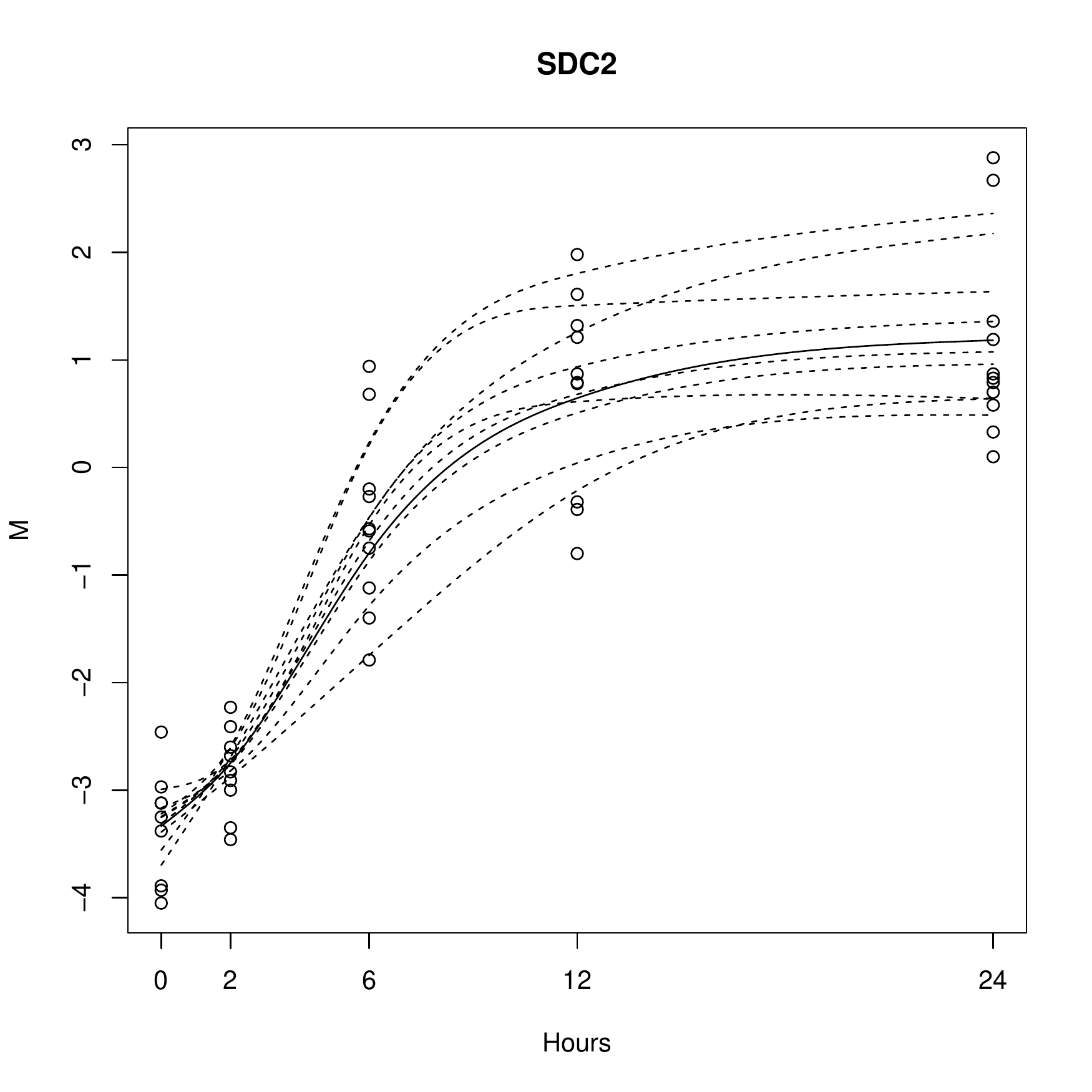}
\caption{\label{fig:sdc2}Fit to an example transcript from the BCG genomics data set obtained under
the
skew-\textit{t}-normal multi-level reduced-rank FPCA model. This transcript, corresponding to gene
SDC2, was found to have the highest positive loading on the third principal component function given
in Figure 5 in the main text. This principal component function account for those probes which are
induced or repressed slowly until around $8$ hours before levelling off. Unsurprisingly, SDC2 is a
prime example of this.}
\end{figure}

\begin{landscape}
\begin{table}[p]
\centering
\begin{tabular}{ccccc}
& & \multicolumn{3}{c}{Number of replicates} \\
& & $5$ & $10$ & $20$ \\
& $100$ & $0.000406$ ($0.000424$) & $0.000195$ ($0.000205$) & $0.0000953$ ($0.0000999$) \\
Number of variables & $1000$ & $0.000348$ ($0.000388$) & $0.000167$ ($0.000186$) & $0.0000813$
($0.0000903$) \\
& $10000$ & $0.000342$ ($0.000387$) & $0.000165$ ($0.000185$) & $0.0000799$ ($0.0000897$)
\end{tabular}
\caption[Estimation error for variable-level curves using the Gaussian
multi-level reduced-rank fPCA model]
{\label{table:multi-level fPCA table}Estimation error for variable-level curves using the Gaussian
multi-level reduced-rank fPCA model. Values shown are the mean (standard deviation) MSE across all
variables in $1000$ simulated data sets}
\end{table}

\begin{table}[p]
\centering
\begin{tabular}{ccccc}
& & \multicolumn{3}{c}{Number of replicates} \\
& & $5$ & $10$ & $20$ \\
& $100$ & $0.00226$ ($0.00224$) & $0.00113$ ($0.00112$) & $0.000562$ ($0.000559$) \\
Number of variables & $1000$ & $0.00225$ ($0.00223$) & $0.00113$ ($0.00112$) & $0.000564$
(0.000561) \\
& $10000$ & $0.00226$ ($0.00224$) & $0.00113$ ($0.00112$) & $0.000564$ ($0.000560$)
\end{tabular}
\caption[Estimation error for variable-level curves using the Gaussian
single-level reduced-rank fPCA model]
{\label{table:single-level fPCA table}Estimation error for variable-level curves using the Gaussian
single-level reduced-rank fPCA model. Values shown are the mean (standard deviation) MSE across all
variables in $1000$ simulated data sets}
\end{table}
\end{landscape}

\bibliographystyle{../myplainnat}
\bibliography{../../References/references}